\documentclass{article}
\usepackage{ijcai23}

\usepackage{times}
\usepackage{soul}
\usepackage{url}
\usepackage[hidelinks]{hyperref}
\usepackage[utf8]{inputenc}
\usepackage[small]{caption}
\usepackage{graphicx}
\usepackage{amsmath}
\usepackage{amssymb}
\usepackage{amsthm}
\usepackage{booktabs}
\usepackage[linesnumbered, ruled]{algorithm2e}
\usepackage[switch]{lineno}
\usepackage{xcolor}
\usepackage{balance}
\usepackage{multirow}
\usepackage{graphicx}
\usepackage{subcaption}
\usepackage{subfiles}
\usepackage{marvosym}

\title{stMCDI: Masked Conditional Diffusion Model with Graph Neural Network for Spatial Transcriptomics Data Imputation}

\author{
Xiaoyu Li$^1$
\and
Wenwen Min$^1$\textsuperscript{(\Letter)}
\and 
Shunfang Wang$^1$
\and
Changmiao Wang$^2$
\and
Taosheng Xu$^3$
\affiliations
$^1$School of Information Science and Engineering, Yunnan University\\
$^2$Shenzhen Research Institute of Big Data\\
$^3$Hefei Institutes of Physical Science, Chinese Academy of Science
\emails Correspondence author: minwenwen@ynu.edu.cn
}

\begin{document}
\maketitle

\begin{abstract}
Spatially resolved transcriptomics represents a significant advancement in single-cell analysis by offering both gene expression data and their corresponding physical locations. 
However, this high degree of spatial resolution entails a drawback, as the resulting spatial transcriptomic data at the cellular level is notably plagued by a high incidence of missing values.
Furthermore, most existing imputation methods either overlook the spatial information between spots or compromise the overall gene expression data distribution.
To address these challenges, our primary focus is on effectively utilizing the spatial location information within spatial transcriptomic data to impute missing values, while preserving the overall data distribution.
We introduce \textbf{stMCDI}, a novel conditional diffusion model for spatial transcriptomics data imputation, which employs a denoising network trained using randomly masked data portions as guidance, with the unmasked data serving as conditions. 
Additionally, it utilizes a GNN encoder to integrate the spatial position information, thereby enhancing model performance.
The results obtained from spatial transcriptomics datasets elucidate the performance of our methods relative to existing approaches.
\end{abstract}

\section{Introduction}
Spatial transcriptomics has revolutionized our understanding of cellular communication and functionality within specific microenvironments. 
This technology facilitates the precise localization of cells within tissues, crucial for elucidating tissue structure and function, and has significantly advanced single-cell biomics analysis \cite{sc1,sc2}. 
However, the high resolution inherent in spatial transcriptomics also introduces certain limitations \cite{st1}.
The low RNA capture rate of this sequencing method potentially limits its ability to detect a wide range of genes, often resulting in data with numerous missing values.
Consequently, researchers have developed imputation methods to accurately estimate missing data values, thereby enhancing data quality \cite{pre-imputation1,pre-imputation2,pre-imputation3}.
In recent years, there has been a significant surge in the development of imputation techniques for spatial transcriptomics data. 
These techniques primarily encompass generative probabilistic models, matrix factorization, and deep learning models \cite{DCA,saver,sti1}.

Advancements in the field of missing data imputation are particularly notable in computer vision and natural language processing. 
However, accurately imputing spatial transcriptomics data continues to pose significant challenges.
Traditional methods are inadequate, as they frequently overlook crucial spatial information, merely reconstructing gene expression matrices \cite{sti2}. 
To address this issue, recent research has shifted towards a graph-based approach \cite{scgnn,graphcpg}, leveraging spatial location information in spatial transcriptomics data. 
This methodology \cite{pre-imputation4,pre-imputation5} constructs a graph using spatial location data and employs a graph neural network for processing. 
While this approach incorporates spatial information, it can potentially distort data distribution, leading to suboptimal imputation results. 
Furthermore, the absence of true labels for spatial transcriptomics data complicates the assessment of imputation accuracy.
In the fields of computer vision and natural language processing, self-supervised learning is frequently employed to compensate for the lack of real labels. 
Consequently, developing an effective imputation method to manage missing spatial transcriptomics data represents a critical challenge that requires resolution.

In this paper, we introduce a novel method for spatial transcriptomics data imputation, \textbf{stMCDI}.
Our method leverages spatial position information to construct a graph about the spot. 
This graph is then used by a graph neural network to learn a latent representation, incorporating spatial position information. 
To address the challenge of working without labels, our approach is inspired by techniques from related research \cite{mae,graphmae}. 
We mask a portion of the original data values, rendering them invisible to the GNN encoder.
This masking step is crucial for our training process. 
We then apply a re-masking technique to the latent representation and employ a conditional score-based diffusion model for training the denoising network. 
The unmasked part of the data serves as a prior condition during this training phase. Additionally, the re-masked part undergoes processing by our training diffusion model. 

We highlight the main contribution as follows:
\begin{itemize}
    \item We employ a graph encoder to integrate gene expression matrices with spatial location information from spatial transcriptomic data.

    \item Employing a masked strategy, our model can predict unknown segments based on known data segments, thereby enhancing imputation performance. Simultaneously, it functions as a self-supervised learning method, providing corresponding labels for the model.

    \item Utilizing the conditional diffusion model, the known segment of the data is incorporated as a priori conditions into the diffusion model, thereby enhancing the model's ability to align with the data's distribution and improving imputation performance.
\end{itemize}

\section{Related Work}
\subsection{Self-supervised Learning for Data Imputation}
Self-supervised learning, which effectively utilizes large amounts of unlabeled data, has gained tremendous attention across various fields. 
In NLP field, for instance, self-supervised models have shown remarkable capability in learning rich data representations without relying on labeled data \cite{bert,gpt3}. 
In time series imputation and prediction, this method involves masking or modifying data segments. Models are trained to predict or reconstruct this missing data, thereby learning the underlying data distribution. \cite{tsi1,tsi2,tsi3}. 
In image processing, self-supervised learning is vital for tasks like denoising and repairing images. A typical method involves masking parts of an image, with the model learning to fill these gaps, thereby improving its understanding of spatial context and image representation. \cite{image1,image2}.

\subsection{Imputation Data via Generative Model}
Deep generative models, such as GANs and VAEs, are increasingly central in research for imputing missing data.
The GAIN framework, a novel approach utilizing adversarial training, capitalizes on the capability of GANs to generate realistic data distributions \cite{GAIN}. 
This enables GAIN to effectively impute missing data, ensuring consistency with the observed data. 
Similarly, VAEs, with their probabilistic foundations and ability to handle complex distributions, are adept at learning the data's underlying representation, making them vital for data imputation \cite{VAE}. 
Utilizing generative models for data imputation poses challenges, including potential bias in cases of extensive missing data and interpretability issues due to their complexity, especially in sensitive areas like medical data handling \cite{pre-gen-imputation1,pre-gen-imputation2}.

\subsection{Spatial Transcriptomics Data Imputation}
Spatial transcriptomics technology, known for providing both RNA expression patterns and their spatial information, has garnered significant interest \cite{pre-st-im1,pre-st-im2}. 
Despite the rapid development of improved techniques in this field, those that aim to offer comprehensive spatial transcriptome-level profiles face challenges, notably dropout problems stemming from high rates of missing data. Addressing this, the strategy of data imputation has emerged as a key method to mitigate these technical issues.
There are currently several common imputation methods, such as SpaGE \cite{SpaGE}, stPlus \cite{stPlus}, gimVI \cite{gimVI}, Tangram \cite{Tangram}.
However, the performance of several common methods was lower than expected, indicating that there is still a gap in imputation tools for handling missing events in spatially resolved transcriptomics.
\section{Proposed Method: stMCDI}
\begin{figure*}[h]%
\centering
\includegraphics[width=0.9\textwidth]{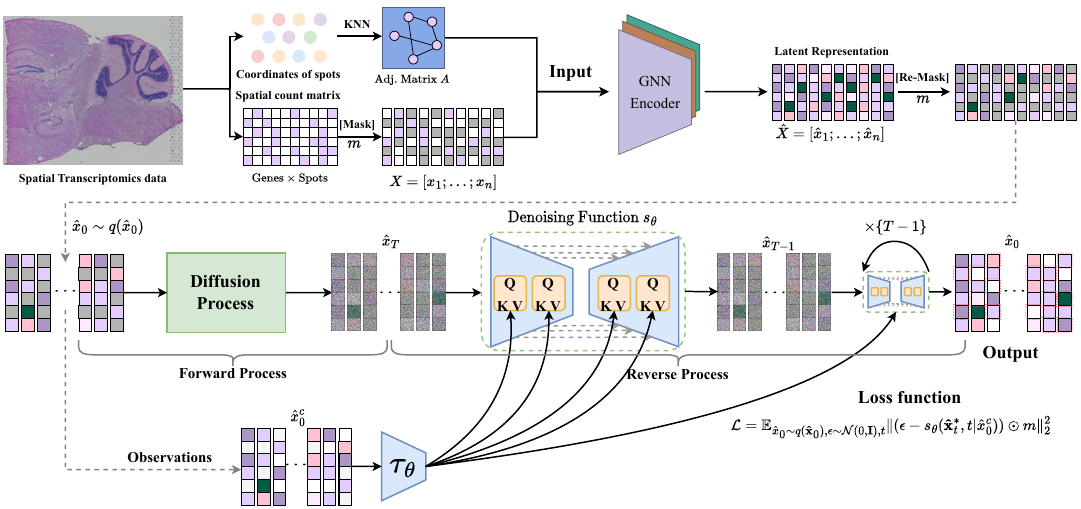}
\caption{The network architecture of the proposed stMCDI model. Our model input has two parts: spot gene expression matrix and spot spatial location information. Build a graph based on the location information of each adjacent spot. Then the diffusion model is used to restore the masked representation to achieve the purpose of imputation.}
\label{fig1}
\end{figure*}

We propose a conditional diffusion model, stMCDI, for spatial transcriptomic data imputation (Figure \ref{fig1}). The model is able to perform accurate imputation using the spatial position information of the idling data as well as the information available in the observations.

\subsection{Problem formulation}
In our problem, we work with two types of input data: the gene expression matrix \(X = \{ x_i\}_{i}^{n}\in \mathbb{R}^{n \times p}\) and the spatial location information of spots \(C \in \mathbb{R}^{n \times 2}\).
Here, \(n\) represents the number of spots and \(p\) represents the number of gene features. 
In the gene expression matrix $X$, some gene expressions are incomplete due to limitations in sequencing technology. 
To address this, we define \(X\) as \(X = X^c \cup X^*\), where \(X^c\) represents the known part of the data and \(X^*\) represents the missing part. 
Our objective is to develop an imputation function \(\mathcal{F}: X \rightarrow \bar{X}\), which transforms \(X\) with its missing values into a complete matrix \(\bar{X}\) without any missing values. 
This function, \(\mathcal{F}\), aims to accurately estimate and replace the missing values.
In contrast to traditional diffusion model applications where vast amounts of data are typically available, our research field often deals with limited datasets. 
Consequently, we treat each spot as an individual sample. 
The strategy is to predict the missing parts of each spot by understanding and utilizing the unique data distribution characteristics of each spot.

To evaluate the performance of $\mathcal{F}$, we use four key evaluation metrics: Pearson Correlation Coefficient (PCC), Cosine Similarity (CS), Root Mean Square Error (RMSE), and Mean Absolute Error (MAE). 
For more details about the evaluation indicators, please refer to the Appendix D.

\subsection{Mask and re-mask strategy in stMCDI}
Since spatial transcriptomic data lacks labels, we adopt a self-supervised learning approach. 
This method involves initially masking a portion of the original data for two key purposes:
\textbf{i)} The masked part of the data serves as a pseudo \textit{label} for subsequent use, allowing for the calculation of various evaluation indicators.
\textbf{ii)} Masking a portion of the data enables the GNN encoder to better fit the deep distribution and semantic structure during the training process, as the model learns to predict the missing information.
Inspired by how large language models comprehend context, we apply a re-masking technique to the latent representation reconstructed by the GNN encoder. 
This is done for two primary reasons:
\textbf{i)} To improve the model's ability to predict data gradient distribution.
\textbf{ii)} To use the unmasked part as a prior condition, guiding the diffusion model in restoring the masked portion and thus enhancing the model's overall performance.

\subsection{GNN Encoder in stMCDI for integrating ST location information}
\subsubsection{Constructing adjacency matrices}
Since spatial transcriptomic data lacks inherent graph structure, we manually construct an adjacency matrix to better utilize its spatial location information. 
In this matrix, the spatial position of each spot is represented in two dimensions. 
To determine the connections between spots, we calculate the Euclidean distance between each pair of spots, \( p_i = (x_i,y_i) \) and \( p_j = (x_j, y_j) \). 
The distance \( d \) between them is given by:
\begin{equation}
    d_{ij} = \sqrt{(x_i - x_j)^2 + (y_i - y_j)^2}.
\end{equation}
This distance calculation helps us in constructing the adjacency matrix. 
We establish adjacency edges between each spot and its five closest neighbors, resulting in the adjacency matrix \( \mathbf{A} \). In \( \mathbf{A} \), the elements \( \mathbf{A}_{ij} \) are defined as follows:
\begin{equation}
    A_{ij} =
        \begin{cases}
            1, & \text{if there is an edge between } \text{spot}~i \text{ and } j\\
            0, & \text{otherwise}
        \end{cases}
\end{equation}

\subsubsection{Constructing latent representation with GCN}
In spatial transcriptomic data analysis, effectively utilizing spatial location information is crucial. 
To achieve this, we implement a graph convolutional network (GCN) as an encoder. 
The GCN encoder integrates information from neighboring spots. 
This ensures that the gene expression characteristics of each spot are not considered in isolation but are blended with those of its adjacent neighbors. 
Through this integration, we achieve multi-scale fusion of gene expression data, thereby enhancing the richness and semantic depth of gene expression information.
As a result, it produces an enriched overall representation of the data, characterized by enhanced informational content and greater biological significance. 
The building of latent representations in the GCN encoder is defined as follows:
\begin{equation}
    \begin{split}
        & \hat{X} = GCN(X, \mathbf{A}), \\
        & \hat{X} = (\hat{x}_1, \dots, \hat{x}_n),~X = (x_1, \dots, x_n).
    \end{split}
\end{equation}
Among them, $\mathbf{X}$ represents the gene expression matrix, $\mathbf{\hat{X}}$ represents the reconstructed latent representation matrix, and $\mathbf{A}$ is the constructed adjacency matrix.

\subsection{Conditional score-based diffusion model in stMCDI}
\subsubsection{Denoising diffusion probabilistic model and Score-based diffusion model} \label{ddpm}
In the realm of denoising diffusion probabilistic models \cite{ddpm2015,ddpm2020}, consider the task of learning a model distribution \( p_\theta(\mathbf{x}_0) \) that closely approximates a given data distribution \( q(\mathbf{x}_0) \). 
Suppose we have a sequence of latent variables \( \mathbf{x}_t \) for \( t = 1, \ldots, T \), existing within the same sample space as \( \mathbf{x}_0 \), which is denoted as \( \mathcal{X} \). 
DDPMs are latent variable models that are composed of two primary processes: the forward process and the reverse process. 
The forward process is defined by a Markov chain, described as follows:
\begin{equation}
\begin{split}
    &q(\mathbf{x}_{1:T} | \mathbf{x}_0) := \prod_{t=1}^{T} q(\mathbf{x}_t | \mathbf{x}_{t-1}),  \\
    \text{where } &q(\mathbf{x}_t | \mathbf{x}_{t-1}) := \mathcal{N}(\sqrt{1 - \beta_t}\mathbf{x}_{t-1},  \beta_t\mathbf{I}).
\end{split}
\label{eq_ddpm_forward}
\end{equation}
and the variable \(\beta_t\) is a small positive constant indicative of a noise level. The sampling of \(x_t\) can be described by the closed-form expression \(q(x_t | x_0) = \mathcal{N}(x_t; \sqrt{\alpha_t}x_0, (1 - \alpha_t)\mathbf{I})\), where \(\hat{\alpha}_t := 1 - \beta_t\) and \(\alpha_t\) is the cumulative product \(\alpha_t := \prod_{i=1}^t \hat{\alpha}_i\). Consequently, \(x_t\) is given by the equation \(x_t = \sqrt{\alpha_t}x_0 + (1 - \alpha_t)\epsilon\), with \(\epsilon \sim \mathcal{N}(0, \mathbf{I})\). In contrast, the reverse process aims to denoise \(x_t\) to retrieve \(x_0\), a process which is characterized by the ensuing Markov chain:
\begin{equation}
\begin{aligned}
&p_\theta(\mathbf{x}_{0:T}) := p(\mathbf{x}_T) \prod_{t=1}^{T} p_\theta(\mathbf{x}_{t-1} | \mathbf{x}_t), \quad \mathbf{x}_T \sim \mathcal{N}(0, \mathbf{I}), \\
&p_\theta(\mathbf{x}_{t-1} | \mathbf{x}_t) := \mathcal{N}(\mathbf{x}_{t-1}; \mu_\theta(\mathbf{x}_t, t), \sigma^2_\theta(\mathbf{x}_t, t)\mathbf{I}),  \\
& \mu_\theta(\mathbf{x}_t, t) = \frac{1}{\alpha_t} \left( \mathbf{x}_t - \frac{\beta_t}{\sqrt{1-\alpha_t}} \epsilon_\theta(\mathbf{x}_t, t) \right), \\
& \sigma_\theta(\mathbf{x}_t, t) = \beta_t^{1/2}, \\
& \text{where } \beta_t = 
\begin{cases} 
\frac{1-\bar{\alpha}_{t-1}}{1-\bar{\alpha}_t} \beta_1, & \text{for } t > 1, \\
\beta_1, & \text{for } t = 1,
\end{cases} \\
& \text{and } \epsilon_\theta(\mathbf{x}_t, t) \text{ is a trainable denoising function.}
\end{aligned}
\label{eq_ddpm_reverse}
\end{equation}

In our proposed method, we use a score-based diffusion model, whose main difference from the traditional DDPM is the different coefficients in the sampling stage. 
Due to space limitations, specific derivation can be found in Appendix~A.1.

\subsubsection{Conditioning mechanisms in stMCDI}
The method we propose aims to estimate the entire dataset by using the known data as a priori conditions. 
This approach is designed to effectively achieve the purpose of data imputation in spatial transcriptomic analysis. 
By leveraging the existing known data, our model can infer the missing values, thereby reconstructing a complete and accurate representation of the entire dataset.
We denote the observed (unmasked) data as $x_{j, 0}^c = m \odot x_{j, 0}$, where $x_{j,0}$ mean the $j^{\text{th}}$ sample in 0 step, and $\odot$ denotes matrix element-wise multiplication and 
$m \in \{ 0, 1 \}^{n \times 1}$ is the elemen-wise indicator with ones be observed and zeros be masked.
So, our goal is to estimate the posterior $p((E_{n,1} - m) \odot x | M \odot x)$, where $E_{n, 1}$ is all-ones matrix of dimensions $n\times 1$.
We also denote the imputation data as $x_{j, t}^{*}$, where t is time step.
Therefore, our conditional mechanisms in stMCDI's objective is to estimate the probabilistic:
\begin{equation}
    p_\theta(x_{j,t-1}^{*} | x_{j, t}^{*}, x_{j, 0}^c).
\end{equation}
In order to better use the observed values as a priori conditions for the diffusion model to perform missing value imputation, we transform the Eq. (\ref{eq_ddpm_forward}) and Eq. (\ref{eq_ddpm_reverse}) into:
\begin{equation}
\small
\begin{split}
&p_\theta(x_{j, 0:T}^{*} | x_{j, 0}^c) := p(x_{j, T}^{*}) \prod_{t=1}^{T} p_\theta(x_{j, t-1}^{*} | x_{j, t}^{*}, x_{j, 0}^c), x_{j, T}^{*} \sim \mathcal{N}(0, \mathbf{I}). \\
&p_\theta(x_{j, t-1}^{*} | x_{j, t}^{*}, x_{j, 0}^c) := \mathcal{N}(x_{j, t-1}^{*}; \mu_\theta(x_{j, t}^{*}, t | x_{j,0}^c), \sigma_\theta(x_{j,t}^{*}, t | x_{j,0}^c)\mathbf{I}).
\end{split}
\label{eq_mcdi_forward_reverse}
\end{equation}
We can optimize the Eq. (\ref{eq_mcdi_forward_reverse}) parameters by minimizing the variational lower bound:
\begin{equation}
\small
\mathbb{E}_{q}\left[ -\log p_{\theta}(x_{j, 0} \mid x_{j, 0}^c) \right] \leq \mathbb{E}_{q}\left[ -\log \frac{p_{\theta}(x_{j, 0:T} \mid x_{j, 0}^c)}{q(x_{j, 1:T} \mid x_{j, 0})} \right].
\label{elbo_conditon}
\end{equation}
Also we can get a simplified training objective:
\begin{equation}
     \mathbb{E}_{x_{j, 0}\sim q(\mathbf{x}_{j, 0}), \epsilon\sim\mathcal{N}(0,\mathbf{I}),t}\|(\epsilon - s_{\theta}(\mathbf{x}_{j, t}^*, t | x_{j, 0}^c))\|^2_2 .
    \label{eq11}
\end{equation}

The goal of conditioner $\tau_{\theta}$ is a multilayer perceptron (MLP) that converts unmasked part of data as an input condition $\tau_{\theta}(x_{j, 0}^c)$.
Inspired by other generative models, the diffusion model essentially models a conditional distribution of the form $p(z|y)$, where $y$ is a prior condition. 
This can be achieved by training a conditional denoising function $s_{\theta}(x_{j,t}, t, y)$, and controlling the generation of the final target by inputting the condition $y$.
We enhance the underlying UNet's Backbone through the cross-attention mechanism, turning SDM into a more flexible conditional diffusion model, allowing us to better learn the gradient distribution of data, thereby restoring a data distribution without \textbf{\textit{missing}}.
We introduce a specific encoder $\tau_{\theta}$, which projects the representation except the mask part to the intermediate representation $\tau_\theta(x_{j, 0}^c) $, and then maps it to the intermediate layers of the UNet through the cross attention layer:
\begin{equation}
    \begin{split}
        &\text{Attenntion}(Q,K,V)=\text{softmax}\left( \frac{QK^T}{\sqrt{d}}\right) \cdot V, \text{with} \\
        &Q = W_Q^{(i)} \cdot \varphi_i(x_{j, T}^{*}), K=W_K^{(i)} \cdot \tau_\theta(x_{j, 0}^c), V=W_V^{(i)} \cdot \tau_\theta(x_{j, 0}^c)
    \end{split}
\end{equation}
In this place equation, $\varphi_i(x_{j, T}^{*})$ denotes a intermediate representation of the UNet implementing $\epsilon_\theta$ and $W_Q^{(i)}, W_K^{(i)}, W_V^{(i)}$ are learnable projection metrices.

\subsubsection{Imputation with stMCDI}
We focus on refining the conditional diffusion model characterized by the reverse process described in Eq.~(\ref{eq_mcdi_forward_reverse}). 
Our goal is to accurately model the conditional distribution $p\left(x_{j, t-1}^{*} | x_{j, t}^{*}, x_{j, 0}^{c}\right)$ without resorting to approximations. 
To achieve this, we adapt the parameterization of DDPM from Eq. (\ref{eq_ddpm_reverse}) for the conditional setting. 
We introduce a conditional denoising function $s_{\theta} : \left(\mathcal{X}^* \times \mathbb{R} \mid \mathcal{X}^c\right) \rightarrow \mathcal{X}^*$ that accepts the conditional observations $x_{j, 0}^c$ as input parameters.
Building on this, we employ the parameterization with $\epsilon_{\theta}$ as follows:
\begin{equation}
\begin{split}
\mu_{\theta}(x_{j, t}^*,t | x_{j, 0}^c) &= \mu^{\text{Score-based}} \left(x_{j, t}^{*}, t, s_{\theta}\left(x_{j, t}^{*}, t | x_{j, 0}^c\right)\right),\\
\sigma_{\theta}(x_{j, t}^*, t | x_{j, 0}^c) &= \sigma^{\text{Score-based}} \left(x_{j, t}^*, t\right),
\end{split}
\label{eq10}
\end{equation}
where $\mu^{\text{Score-based}}$ and $\sigma^{\text{Score-based}}$ are defined in Section \ref{ddpm}.
Utilizing the function $\epsilon_{\theta}$ and the data $x_0$, we can simulate samples of $x_{j, 0}^*$ by employing the reverse process outlined in Eq. (\ref{eq_mcdi_forward_reverse}). 
During sampling, we treat the known values of $x_{j, 0}$ as the conditional observations $x_{j, 0}^c$ and the unknown values as the targets for imputation $x_{j, 0}^*$. 

\subsection{Loss function of stMCDI}
Specifically, in the presence of conditional observations $x_{j, 0}^c$ and imputation targets $x_{j, 0}^*$, we generate noisy targets $x_{j, t}^* = \sqrt{\alpha_t} x_{j, 0}^* + \sqrt{(1-\alpha_t)}\epsilon$, and proceed to refine $s_\theta$ by minimizing the ensuing loss function:
\begin{equation}
     \mathcal{L} =  \mathbb{E}_{x_{j, 0}\sim q(\mathbf{x}_{j, 0}), \epsilon\sim\mathcal{N}(0,\mathbf{I}),t}\|(\epsilon - s_{\theta}(\mathbf{x}_{j, t}^*, t | x_{j, 0}^c)) \odot M\|^2_2 .
    \label{eq_loss_DM}
\end{equation}
The stMCDI algorithm framework can be found in the Appendix~B.

\section{Experiment}
\subsection{Data sources and data preprocessing}
We compared the performance of our model with other baseline methods on 6 real-world spatial transcriptomic datasets from several representative sequencing platforms. The real-world spatial transcriptomics datasets, including: Mouse Olfactory Bulb (MOB), Human Breast Cancer (HBC), Human Prostate (HP), Human Osteosarcoma (HO), Mouse Liver (ML), Mouse Kidney (MK), used in our experiments were derived from recently published papers on spatial transcriptomics experiments, and our data preprocessing method is consistent with this paper \cite{stBenchmark}.
All datasets come from different species, including mice and humans, and different organs, such as liver and kidneys. Specifically, the number of spots ranges from 278 to 6000, and the gene range ranges from 14192 to 28601.
Detailed information about the dataset can be found in the Appendix~E.
\begin{table*}[htbp]
\centering
\resizebox{0.95\textwidth}{!}{%
\begin{tabular}{l|cccc|cccc|cccc}
\hline
\multirow{2}{*}{\textbf{Method}} & \multicolumn{4}{c|}{\textbf{Dataset\_1: Mouse Olfactory Bulb~(MOB)}}                                           & \multicolumn{4}{c|}{\textbf{Dataset\_2: Human Breast Cancer~(HBC)}}   & \multicolumn{4}{c}{\textbf{Dataset\_3: Human Prostate~(HP)}}                                                  \\ \cline{2-13} 
                        & PCC$\uparrow$                   & Cosine$\uparrow$                & RMSE$\downarrow$                  & MAE$\downarrow$        & PCC$\uparrow$                   & Cosine$\uparrow$                & RMSE$\downarrow$                  & MAE$\downarrow$        & PCC$\uparrow$                   & Cosine$\uparrow$                & RMSE$\downarrow$                  & MAE$\downarrow$                   \\ \hline
Mean                    & -0.005±0              & 0.037±0               & 1.334±0               & 1.013±0               & -0.223±0              & -0.476±0              & 0.973±0               & 0.884±0               & -0.632±0              & -0.237±0              & 1.731±0               & 1.603±0               \\
KNN                     & -0.063±0.0012         & 0.094±0.0018          & 1.103±0.0357          & 0.953±0.0027          & -0.371±0.0001         & 0.148±0.0001          & 0.812±0.0002          & 0.741±0.0001          & -0.0034±0.0017        & 0.223±0.0022          & 0.971±0.0005          & 0.843±0.0007          \\
gimVI                   & 0.513±0.0013          & 0.477±0.0011          & 0.346±0.0022          & 0.241±0.0005          & 0.538±0.0006          & 0.461±0.0006          & 0.460±0.0004          & 0.406±0.0012          & 0.615±0.0002          & 0.619±0.0007          & 0.461±0.0019          & 0.437±0.0022          \\
SpaGE                   & 0.436±0.0026          & 0.503±0.0007          & 0.289±0.0003          & 0.256±0.0012          & 0.578±0.0004          & 0.513±0.0004          & 0.413±0.0011          & 0.389±0.0007          & 0.634±0.0013          & 0.603±0.0024          & 0.423±0.0021          & 0.488±0.0007          \\
Tangram                 & 0.603±0.0002          & 0.526±0.0004          & 0.203±0.0002          & 0.301±0.00021         & 0.603±0.0014          & 0.525±0.0001          & 0.396±0.0021          & 0.351±0.0002          & 0.657±0.0022          & 0.611±0.0017          & 0.401±0.0005          & 0.396±0.0006          \\
stPlus                  & 0.611±0.0003          & 0.518±0.0002          & 0.301±0.0004          & 0.212±0.0016          & 0.651±0.0002          & 0.534±0.0004          & 0.387±0.0012          & 0.404±0.0005          & 0.587±0.001           & 0.663±0.0021          & 0.469±0.0007          & 0.357±0.0221          \\
STAGATE                 & 0.626±0.0021          & 0.536±0.0034          & 0.242±0.0014          & 0.196±0.0026          & 0.626±0.0004          & 0.503±0.0002          & 0.401±0.0002          & 0.363±0.0004          & 0.626±0.0002          & 0.653±0.0002          & 0.403±0.0017          & 0.326±0.0004          \\
DCA                     & 0.465±0.0029          & 0.491±0.0047          & 0.338±0.0009          & 0.289±0.0013          & 0.460±0.0009          & 0.437±0.0011          & 0.483±0.0024          & 0.461±0.0024          & 0.559±0.0003          & 0.572±0.0006          & 0.509±0.0004          & 0.496±0.0004          \\
GraphSCI                & 0.408±0.0026          & 0.509±0.0006          & 0.355±0.0009          & 0.329±0.0006          & 0.481±0.0006          & 0.437±0.0002          & 0.504±0.0005          & 0.476±0.0011          & 0.565±0.0014          & 0.551±0.0003          & 0.479±0.0003          & 0.453±0.0005          \\
SpaFormer               & 0.385±0.0013          & 0.336±0.0016          & 0.505±0.0077          & 0.343±0.0004          & 0.375±0.0002          & 0.381±0.0004          & 0.632±0.0009          & 0.515±0.0003          & 0.537±0.0031          & 0.453±0.0006          & 0.607±0.0028          & 0.598±0.0025          \\
CpG                     & 0.627±0.0017          & 0.511±0.0004          & 0.321±0.0015          & 0.248±0.0004          & 0.475±0.0003          & 0.440±0.0009          & 0.460±0.0001          & 0.413±0.0001          & 0.589±0.0014          & 0.445±0.0061          & 0.587±0.0024          & 0.563±0.0019          \\
GraphCpG                & 0.557±0.0003          & 0.539±0.0009          & 0.288±0.0002          & 0.243±0.0003          & 0.501±0.0001          & 0.443±0.0017          & 0.436±0.0016          & 0.403±0.0022          & 0.629±0.0003          & 0.459±0.0041          & 0.531±0.0013          & 0.518±0.0005          \\
scGNN                   & 0.548±0.0005          & 0.532±0.0007          & 0.307±0.0009          & 0.244±0.0011          & 0.502±0.0015          & 0.455±0.0019          & 0.437±0.0083          & 0.389±0.0015          & 0.602±0.0007          & 0.468±0.0033          & 0.479±0.0007          & 0.502±0.0004          \\
CSDI                    & 0.555±0.0008          & 0.529±0.0008          & 0.293±0.0002          & 0.216±0.0018          & 0.495±0.0004          & 0.481±0.0079          & 0.395±0.0004          & 0.379±0.0006          & 0.624±0.0005          & 0.569±0.0014          & 0.497±0.0051          & 0.484±0.0043          \\
\textbf{stMCDI~(Ours)}    & \textbf{0.645±0.0004} & \textbf{0.557±0.0018} & \textbf{0.188±0.0002} & \textbf{0.147±0.0012} & \textbf{0.703±0.0003} & \textbf{0.623±0.0012} & \textbf{0.304±0.0017} & \textbf{0.332±0.0027} & \textbf{0.714±0.0018} & \textbf{0.726±0.0022} & \textbf{0.367±0.0023} & \textbf{0.316±0.0007} \\ \hline
\multirow{2}{*}{\textbf{Method}} & \multicolumn{4}{c|}{\textbf{Dataset\_4: Human Osteosarcoma~(HO)}}                                             & \multicolumn{4}{c|}{\textbf{Dataset\_5: Mouse Liver~(ML)}}                                                    & \multicolumn{4}{c}{\textbf{Dataset\_6: Mouse Kidney~(MK)}}                                                    \\ \cline{2-13} 
                        & PCC$\uparrow$                   & Cosine$\uparrow$                & RMSE$\downarrow$                  & MAE$\downarrow$                   & PCC$\uparrow$                   & Cosine$\uparrow$                & RMSE$\downarrow$                  & MAE$\downarrow$                   & PCC$\uparrow$                   & Cosine$\uparrow$                & RMSE$\downarrow$                  & MAE$\downarrow$                   \\ \hline
Mean                    & 0.136±0               & -0.469±0              & 1.264±0               & 1.033±0               & 0.047±0               & 0.024±0               & 1.337±0               & 1.454±0               & 0.047±0               & 0.024±0               & 1.337±0               & 1.454±0               \\
KNN                     & 0.157±0.0004          & 0.147±0.0081          & 1.081±0.0081          & 0.958±0.0018          & 0.227±0.0017          & 0.117±0.0041          & 0.947±0.0005          & 1.013±0.0032          & 0.227±0.0017          & 0.117±0.0041          & 0.947±0.0005          & 1.013±0.0032          \\
gimVI                   & 0.454±0.0005          & 0.642±0.0011          & 0.487±0.0011          & 0.436±0.0009          & 0.379±0.0043          & 0.219±0.0055          & 0.412±0.0003          & 0.467±0.0022          & 0.459±0.0004          & 0.344±0.0054          & 0.546±0.0018          & 0.526±0.0026          \\
SpaGE                   & 0.489±0.0005          & 0.613±0.0028          & 0.433±0.0004          & 0.389±0.0004          & 0.315±0.0026          & 0.289±0.0037          & 0.403±0.0079          & 0.496±0.0047          & 0.503±0.0007          & 0.447±0.0001          & 0.567±0.0011          & 0.515±0.0007          \\
Tangram                 & 0.396±0.0127          & 0.637±0.0017          & 0.467±0.0121          & 0.403±0.0002          & 0.403±0.0017          & 0.361±0.0051          & 0.379±0.0021          & 0.422±0.0031          & 0.489±0.0022          & 0.473±0.0003          & 0.476±0.0006          & 0.489±0.0013          \\
stPlus                  & 0.481±0.0056          & 0.625±0.0026          & 0.489±0.0025          & 0.357±0.0001          & 0.511±0.0022          & 0.401±0.0004          & 0.401±0.0007          & 0.378±0.0003          & 0.522±0.0031          & 0.563±0.0013          & 0.516±0.0081          & 0.441±0.0079          \\
STAGATE                 & 0.591±0.0013          & 0.676±0.0027          & 0.466±0.0022          & 0.401±0.0003          & 0.536±0.0013          & 0.489±0.0107          & 0.467±0.0026          & 0.439±0.0011          & 0.679±0.0028          & 0.611±0.0023          & 0.361±0.0002          & 0.379±0.0004          \\
DCA                     & 0.401±0.0032          & 0.571±0.0018          & 0.467±0.0018          & 0.441±0.0014          & 0.488±0.0029          & 0.463±0.0061          & 0.432±0.0079          & 0.504±0.0078          & 0.537±0.0016          & 0.638±0.0084          & 0.413±0.0008          & 0.363±0.0006          \\
GraphSCI                & 0.371±0.0061          & 0.605±0.0002          & 0.543±0.0014          & 0.524±0.0003          & 0.503±0.0011          & 0.537±0.0027          & 0.377±0.0025          & 0.412±0.0034          & 0.567±0.0004          & 0.604±0.0009          & 0.354±0.0004          & 0.323±0.0005          \\
SpaFormer               & 0.374±0.0044          & 0.567±0.0048          & 0.558±0.0008          & 0.538±0.0009          & 0.575±0.0009          & 0.537±0.0016          & 0.411±0.0031          & 0.464±0.0007          & 0.426±0.0007          & 0.513±0.0009          & 0.503±0.0017          & 0.469±0.0053          \\
CpG                     & 0.425±0.0002          & 0.632±0.0004          & 0.564±0.0013          & 0.521±0.0005          & 0.603±0.0003          & 0.571±0.0072          & 0.481±0.0001          & 0.412±0.0009          & 0.517±0.0008          & 0.564±0.0006          & 0.431±0.0053          & 0.412±0.0007          \\
GraphCpG                & 0.431±0.0016          & 0.619±0.0051          & 0.524±0.0031          & 0.558±0.0004          & 0.566±0.0003          & 0.467±0.0004          & 0.437±0.0053          & 0.423±0.0028          & 0.587±0.0056          & 0.729±0.0037          & 0.356±0.0076          & 0.388±0.0009          \\
scGNN                   & 0.431±0.0016          & 0.534±0.0004          & 0.524±0.0031          & 0.534±0.0004          & 0.423±0.0004          & 0.459±0.0001          & 0.397±0.0002          & 0.388±0.0017          & 0.439±0.0027          & 0.693±0.0018          & 0.347±0.0036          & 0.331±0.0051          \\
CSDI                    & 0.519±0.0007          & 0.642±0.0007          & 0.519±0.0003          & 0.478±0.0021          & 0.466±0.0005          & 0.465±0.0008          & 0.423±0.0007          & 0.401±0.0013          & 0.624±0.0015          & 0.743±0.0055          & 0.327±0.0016          & 0.304±0.0029          \\
\textbf{stMCDI~(Ours)}    & \textbf{0.689±0.0007} & \textbf{0.734±0.0028} & \textbf{0.401±0.0022} & \textbf{0.353±0.0013} & \textbf{0.632±0.0047} & \textbf{0.596±0.0004} & \textbf{0.389±0.0124} & \textbf{0.301±0.0004} & \textbf{0.737±0.0004} & \textbf{0.803±0.0019} & \textbf{0.312±0.0027} & \textbf{0.288±0.0042} \\ \hline
\end{tabular}%
}
\caption{Performance comparison between various baselines on the spatial transcriptomic datasets using four evaluation metrics. Among them, the black bold part represents the best performance. It should be noted that KNN means to fill missing values using the nearest spot values. Mean means using the mean value of each spot to fill in missing values.}
\label{tab1}
\end{table*}
\begin{figure*}
    \centering
    \includegraphics[width=0.98\textwidth]{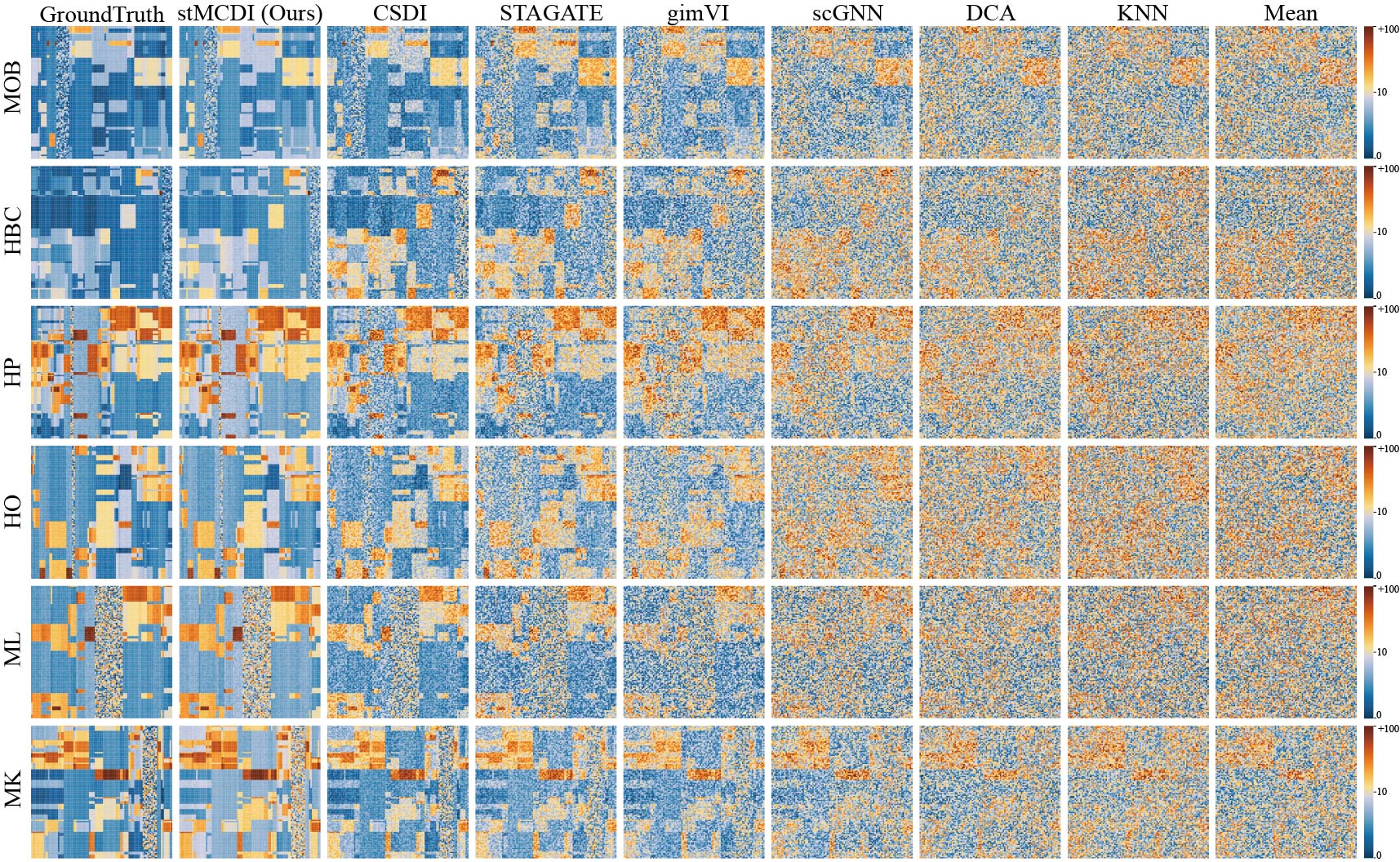}
    \caption{Visualization of the imputation performance of various baseline methods. In this figure, we only show several baselines: CSDI, STAGATE, gimVI, scGNN, DCA, KNN, Mean, etc. For the remaining Baslines, please refer to Appendix~F.2.}
    \label{fig2}
\end{figure*}

\subsection{Baselines}
The performance of stMCDI was first compared with two regular imputation methods, including KNN imputation and mean imputation, as well as several state-of-the-art imputation methods for single-cell data and spatial transcriptomic data.
Finally, we selected fourteen state-of-the-art methods for comparison:
\begin{itemize}
    \item gimVI (\cite{gimVI}):It is a deep generative model for integrating spatial transcriptomics data and scRNA-seq data which can be used to impute missing genes.

    \item SpaGE (\cite{SpaGE}):It is a method that integrates spatial and scRNA-seq datasets to predict whole-transcriptome expressions in their spatial configuration. 
    
    \item Tangram (\cite{Tangram}):It is a method that can map any type of sc/snRNA-seq data, including multimodal data such as those from SHARE-seq, which can be used to reveal spatial patterns of chromatin accessibility.

    \item stPlus (\cite{stPlus}):It is a reference-based method that leverages information in scRNA-seq data to enhance spatial transcriptomics.

    \item STAGATE (\cite{STAGATE}):It is a method that adopts an attention mechanism to adaptively learn the similarity of neighboring spots, and an optional cell type-aware module through integrating the pre-clustering of gene expressions. 

    \item DCA (\cite{DCA}):It is a deep count autoencoder network (DCA) to denoise scRNA-seq datasets. 

    \item GraphSCI (\cite{GraphSCI}):It is an imputation method (GraphSCI) to impute the dropout events in scRNA-seq data based on the graph convolution networks.

    \item SpaFormer (\cite{Spaformer}):It is a transformer-based imputation framework for cellular-level spatial transcriptomic data.
    
    \item CpG (\cite{cpg}):It is a transformer-based imputation method to operate on methylation matrices through combining axial attention with sliding window self-attention. 

    \item GraphCpG (\cite{GraphSCI}):It is a novel graph-based deep learning approach to impute methylation matrices based on locus-aware neighboring subgraphs with locus-aware encoding orienting on one cell type. 

    \item scGNN (\cite{scgnn}):It is a hypothesis-free deep learning framework for scRNA-Seq imputations.

    \item CSDI (\cite{csdi}):It is a novel time series imputation method that utilizes score-based diffusion models conditioned on observed data. 
\end{itemize}

\subsection{Implementation Details}
The hyperparameters of stMCDI are defined as follows: Graph encoder layers to 3 GCN layers; diffusion steps to 2000; learning rate to 6e-6; batch size to 64.
The noise variances were in the range of 10-6 to 0.05.
An exponential moving average (EMA) over model parameters with a rate of 0.926 emploved.
The model is trained on 1 NVIDIA RTX 4090, 24 GB with Adamw optimizer on PyTorch.
Specific details of each parameter can be found in the Appendix ~C.

\subsection{Results}
We carry out multiple experiments on six real-world ST datasets with different the number of spots (Table \ref{tab1}).
The experimental results demonstrate that our proposed method outperforms fourteen baseline methods, achieving the best performance across all four indicators.

To better demonstrate that our proposed method achieves state-of-the-art performance compared to other methods, we use a heatmap of gene expression to visualize the imputation results of stMCDI (as shown in Figure \ref{fig2}). 
We splice the mask part of each sample to form the real label, and we apply the same method to the imputation results of each baseline method. 
To enhance the visualization effect, we perform hierarchical clustering on the spliced mask gene expression matrix, making it exhibit more pronounced pattern-like features, thereby highlighting the effectiveness of the interpolation method.
Due to space limitations, we only show part of the Baseline methods. More results can be found in the Appendix~F.2.

\subsection{Ablation studies}
We conduct ablation experiment to verify the modules of stMCDI in these aspects: 1) Different Mask strategies and different Mask phase. 2) The impact of different Graph Encoder on model performance. 3) Different Mask ratios. 

\subsubsection{Different mask strategy and different mask phase}
In our proposed method stMCDI, we randomly mask the gene expression matrix according to a certain proportion.
We apply this masking process twice: once on the original input data and again on the intermediate latent representation.
To demonstrate the effectiveness of our strategy, we conducted five controlled experiments. These experiments encompass two distinct Mask strategies: \textbf{Mask Spot} and \textbf{Mask Gene}, along with three different Mask phases: \textbf{OR Mask} (masking only the original data), \textbf{LR Mask} (masking only the latent representation), and \textbf{NO Mask}.
The experimental results (Table \ref{tab2}) indicate that our approach is the most effective across six datasets. It also achieves superior performance in four key evaluation metrics compared to other methods.
\begin{table}[htbp]
\centering
\resizebox{0.72\columnwidth}{!}{%
\begin{tabular}{l|cccccc}
\hline
Metrics              & \multicolumn{6}{c}{\textbf{Dataset}}                                                                                                                                                                                                 \\ \hline
\textbf{PCC}$\uparrow$         & \textbf{MOB}                       & \textbf{HBC}                       & \textbf{HP}                        & \textbf{HO}                        & \textbf{ML}                        & \textbf{MK}                        \\ \hline
\textbf{stMCDI~(Ours)} & \textbf{0.645}                     & \textbf{0.703}                     & \textbf{0.714}                     & \textbf{0.689}                     & \textbf{0.632}                     & \textbf{0.737}                     \\
w/ mask Spot         & 0.573                              & 0.603                              & 0.617                              & 0.549                              & 0.581                              & 0.627                              \\
w/ mask Gene         & 0.588                              & 0.612                              & 0.626                              & 0.567                              & 0.593                              & 0.643                              \\
w/o OR mask          & 0.624                              & 0.656                              & 0.677                              & 0.626                              & 0.611                              & 0.639                              \\
w/o LR mask          & 0.602                              & 0.633                              & 0.659                              & 0.587                              & 0.628                              & 0.614                              \\
w/o mask       & 0.569                              & 0.527                              & 0.496                              & 0.533                              & 0.588                              & 0.601                              \\ \hline \hline
\textbf{Cosine}$\uparrow$      & \textbf{MOB}                       & \textbf{HBC}                       & \textbf{HP}                        & \textbf{HO}                        & \textbf{ML}                        & \textbf{MK}                        \\ \hline
\textbf{stMCDI~(Ours)} & \multicolumn{1}{l}{\textbf{0.577}} & \multicolumn{1}{l}{\textbf{0.623}} & \multicolumn{1}{l}{\textbf{0.726}} & \multicolumn{1}{l}{\textbf{0.734}} & \multicolumn{1}{l}{\textbf{0.596}} & \multicolumn{1}{l}{\textbf{0.803}} \\
w/ mask Spot         & \multicolumn{1}{l}{0.511}          & \multicolumn{1}{l}{0.561}          & \multicolumn{1}{l}{0.641}          & \multicolumn{1}{l}{0.681}          & \multicolumn{1}{l}{0.503}          & \multicolumn{1}{l}{0.731}          \\
w/ mask Gene         & \multicolumn{1}{l}{0.507}          & \multicolumn{1}{l}{0.557}          & \multicolumn{1}{l}{0.622}          & \multicolumn{1}{l}{0.673}          & \multicolumn{1}{l}{0.531}          & \multicolumn{1}{l}{0.743}          \\
w/o OR mask          & \multicolumn{1}{l}{0.543}          & \multicolumn{1}{l}{0.489}          & \multicolumn{1}{l}{0.704}          & \multicolumn{1}{l}{0.688}          & \multicolumn{1}{l}{0.567}          & \multicolumn{1}{l}{0.729}          \\
w/o LR mask          & \multicolumn{1}{l}{0.525}          & \multicolumn{1}{l}{0.579}          & \multicolumn{1}{l}{0.693}          & \multicolumn{1}{l}{0.694}          & \multicolumn{1}{l}{0.559}          & \multicolumn{1}{l}{0.759}          \\
w/o mask       & \multicolumn{1}{l}{0.513}          & \multicolumn{1}{l}{0.416}          & \multicolumn{1}{l}{0.589}          & \multicolumn{1}{l}{0.587}          & \multicolumn{1}{l}{0.497}          & \multicolumn{1}{l}{0.721}          \\ \hline \hline
\textbf{RMSE}$\downarrow$        & \textbf{MOB}                       & \textbf{HBC}                       & \textbf{HP}                        & \textbf{HO}                        & \textbf{ML}                        & \textbf{MK}                        \\ \hline
\textbf{stMCDI~(Ours)} & \textbf{0.188}                     & \textbf{0.304}                     & \textbf{0.367}                     & \textbf{0.401}                     & \textbf{0.389}                     & \textbf{0.312}                     \\
w/ mask Spot         & 0.213                              & 0.366                              & 0.427                              & 0.497                              & 0.403                              & 0.379                              \\
w/ mask Gene         & 0.224                              & 0.372                              & 0.438                              & 0.503                              & 0.411                              & 0.406                              \\
w/o OR mask          & 0.215                              & 0.403                              & 0.363                              & 0.459                              & 0.464                              & 0.413                              \\
w/o LR mask          & 0.202                              & 0.358                              & 0.422                              & 0.445                              & 0.451                              & 0.439                              \\
w/o mask       & 0.307                              & 0.402                              & 0.379                              & 0.488                              & 0.426                              & 0.415                              \\ \hline \hline
\textbf{MAE}$\downarrow$         & \textbf{MOB}                       & \textbf{HBC}                       & \textbf{HP}                        & \textbf{HO}                        & \textbf{ML}                        & \textbf{MK}                        \\ \hline
\textbf{stMCDI~(Ours)} & \textbf{0.147}                     & \textbf{0.332}                     & \textbf{0.316}                     & \textbf{0.353}                     & \textbf{0.301}                     & \textbf{0.288}                     \\
w/ mask Spot         & 0.204                              & 0.403                              & 0.411                              & 0.426                              & 0.388                              & 0.369                              \\
w/ mask Gene         & 0.211                              & 0.397                              & 0.423                              & 0.437                              & 0.426                              & 0.403                              \\
w/o OR mask          & 0.164                              & 0.384                              & 0.386                              & 0.439                              & 0.453                              & 0.366                              \\
w/o LR mask          & 0.195                              & 0.362                              & 0.398                              & 0.415                              & 0.437                              & 0.482                              \\
w/o mask       & 0.288                              & 0.401                              & 0.417                              & 0.441                              & 0.449                              & 0.467                              \\ \hline
\end{tabular}%
}
\caption{
Ablation experiments on different masked strategy and different masked phase in stMCDI.
}
\label{tab2}
\end{table}

\begin{table}[htbp]
\centering
\resizebox{0.72\columnwidth}{!}{%
\begin{tabular}{l|cccccc}
\hline
Metrics              & \multicolumn{6}{c}{\textbf{Dataset}}                                                                         \\ \hline
\textbf{PCC}$\uparrow$         & \textbf{MOB}   & \textbf{HBC}   & \textbf{HP}    & \textbf{HO}    & \textbf{ML}    & \textbf{MK}    \\ \hline
\textbf{stMCDI~(Ours)} & \textbf{0.645} & \textbf{0.703} & \textbf{0.714} & \textbf{0.689} & \textbf{0.632} & \textbf{0.737} \\
w/ GAT               & 0.597          & 0.501          & 0.655          & 0.578          & 0.607          & 0.688          \\
w/ GIN               & 0.51           & 0.537          & 0.569          & 0.511          & 0.601          & 0.671          \\
w/ MLP               & 0.505          & 0.517          & 0.526          & 0.488          & 0.579          & 0.665          \\
w/o GNN              & 0.464          & 0.481          & 0.509          & 0.447          & 0.566          & 0.639          \\ \hline \hline
\textbf{Cosine}$\uparrow$      & \textbf{MOB}   & \textbf{HBC}   & \textbf{HP}    & \textbf{HO}    & \textbf{ML}    & \textbf{MK}    \\ \hline
\textbf{stMCDI~(Ours)} & \textbf{0.557} & \textbf{0.623} & \textbf{0.726} & \textbf{0.734} & \textbf{0.596} & \textbf{0.803} \\
w/ GAT               & 0.532          & 0.526          & 0.664          & 0.646          & 0.513          & 0.734          \\
w/ GIN               & 0.514          & 0.507          & 0.613          & 0.598          & 0.501          & 0.705          \\
w/ MLP               & 0.503          & 0.485          & 0.586          & 0.557          & 0.489          & 0.689          \\
w/o GNN              & 0.489          & 0.426          & 0.537          & 0.521          & 0.457          & 0.655          \\ \hline \hline
\textbf{RMSE}$\downarrow$        & \textbf{MOB}   & \textbf{HBC}   & \textbf{HP}    & \textbf{HO}    & \textbf{ML}    & \textbf{MK}    \\ \hline
\textbf{stMCDI~(Ours)} & \textbf{0.188} & \textbf{0.304} & \textbf{0.367} & \textbf{0.401} & \textbf{0.389} & \textbf{0.312} \\
w/ GAT               & 0.308          & 0.398          & 0.411          & 0.452          & 0.402          & 0.374          \\
w/ GIN               & 0.379          & 0.451          & 0.436          & 0.467          & 0.411          & 0.388          \\
w/ MLP               & 0.431          & 0.512          & 0.565          & 0.493          & 0.423          & 0.404          \\
w/o GNN              & 0.447          & 0.569          & 0.587          & 0.537          & 0.436          & 0.476          \\ \hline \hline
\textbf{MAE}$\downarrow$         & \textbf{MOB}   & \textbf{HBC}   & \textbf{HP}    & \textbf{HO}    & \textbf{ML}    & \textbf{MK}    \\ \hline
\textbf{stMCDI~(Ours)} & \textbf{0.147} & \textbf{0.332} & \textbf{0.316} & \textbf{0.353} & \textbf{0.301} & \textbf{0.288} \\
w/ GAT               & 0.188          & 0.367          & 0.413          & 0.413          & 0.359          & 0.326          \\
w/ GIN               & 0.224          & 0.381          & 0.426          & 0.451          & 0.428          & 0.377          \\
w/ MLP               & 0.289          & 0.405          & 0.435          & 0.467          & 0.446          & 0.389          \\
w/o GNN              & 0.346          & 0.455          & 0.473          & 0.503          & 0.461          & 0.429          \\ \hline
\end{tabular}%
}
\caption{
Ablation experiments on different encoders of GNN. We tried different types of graph neural networks and found that GCN was the best graph encoder for our task.}
\label{tab3}
\end{table}

\subsubsection{Graph embedding with different graph neural network}
Our stMCDI method includes a Graph embedding layer to enhance the processing of spatial position information in spatial transcriptomics data. This layer merges spatial data into the gene expression matrix, creating a graph with spatial characteristics. As a result, gene expression's latent representation is augmented with adjacent position details, improving the data representation.
To evaluate the efficiency of GNN in integrating spatial location information, we conducted a series of comparative studies. 
Our experimental results (Table~\ref{tab3}) reveal that the GCN exhibits superior performance in our specific tasks.
However, this finding does not diminish the effectiveness of other GNNs. Instead, it suggests that GCN is particularly well-suited for the specific requirements of our tasks.
\begin{figure}[htbp]
\centering
\includegraphics[width=0.48\textwidth]{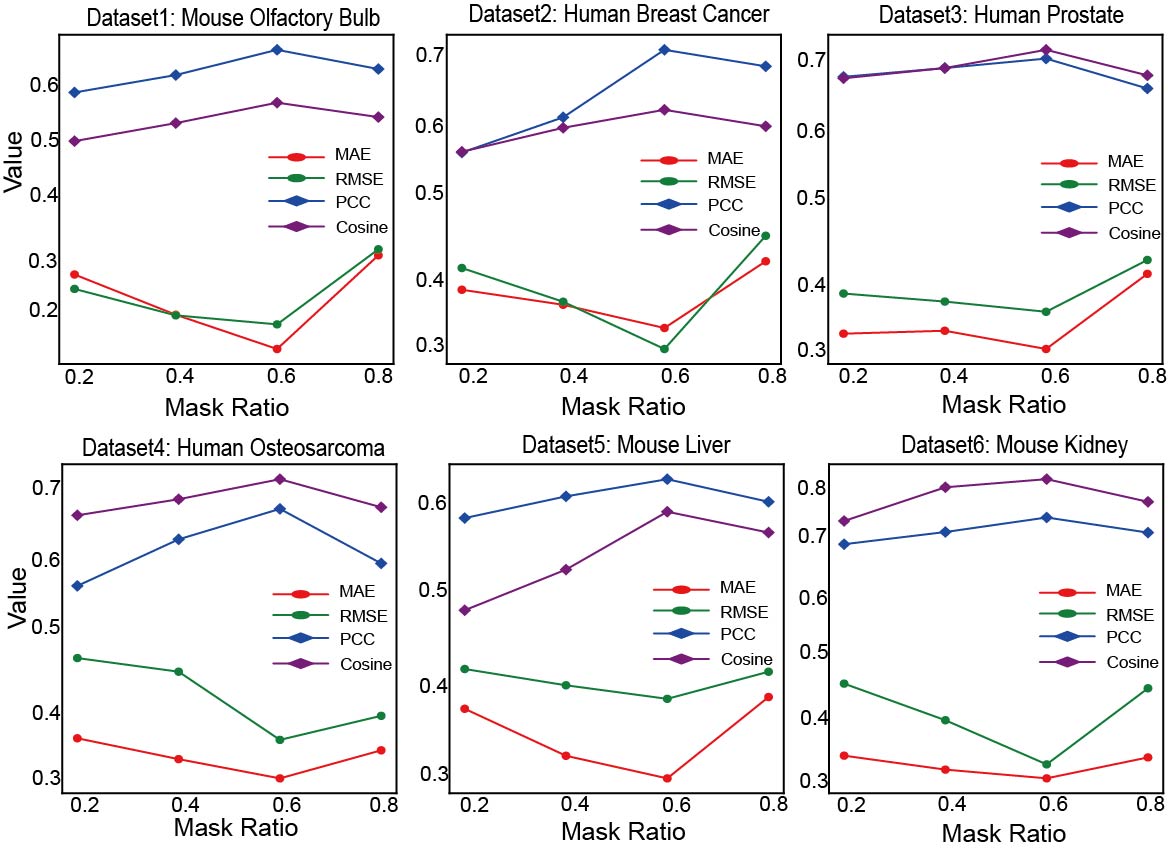} 
\caption{Different mask ratio of imputation performance in four metrics. We adopt different mask proportions for each sample, and we find that the performance of stMCDI reaches its best when the mask proportion is around 60\%.}
\label{fig3}
\end{figure}

\subsubsection{The impact of Mask ratio on model performance}
In our method, the ratio of our mask is 60\%, and we want to explore the impact of different mask ratios on the imputation performance of the model.
We give the comparison of using different masking ratios in Figure \ref{fig3}.
Therefore, we change the parameter mask ratio and observed the changes in model performance under different mask ratios.
We found that as the mask ratio increases, the interpolation performance of the model decreases. The possible reason is that the model does not learn relevant representations from limited data.

\section{Conclusion}
In this paper, we propose a novel method: \textbf{stMCDI}, masked conditional diffusion model with graph neural network for
spatial transcriptomics data imputation.
By masking the latent representation and using the non-mask part as a condition, the diffusion model is well guided and controllable during reverse process, ensuring the quality of imputation.
Experiments on the real spatial transcriptomic datasets show that our proposed method \textbf{stMCDI} achieves the best performance compared with other baselines. 
Moreover, ablation experiments demonstrate the effectiveness and reliability of our method.
Although the \textbf{stMCDI} experimental results are promising, there is still a lot of scope for improvement. 
We can try multi-modal imputation methods, such as introducing single-cell data into the diffusion model as a priori conditions. Secondly, how the imputation data improves the performance of downstream analysis is still a question we consider.

\section*{Acknowledgments}
The work was supported in part by the National Natural Science Foundation of China (62262069), in part by the Yunnan Fundamental Research Projects under Grant (202201AT070469, 202301BF070001-019).

\small
\bibliographystyle{named}
\bibliography{ijcai23}

\appendix
\twocolumn[
\begin{@twocolumnfalse}
     \section*{ \centering{\LARGE  Appendix}} \vspace{15pt}
\end{@twocolumnfalse}
]

\maketitle
\section{Details of denoising diffusion probabilistic models}
\subsection{Overview of DDPM and score-based model} \label{ap_overview}
In the realm of denoising diffusion probabilistic models \cite{ddpm2015,ddpm2020}, consider the task of learning a model distribution \( p_\theta(\mathbf{x}_0) \) that closely approximates a given data distribution \( q(\mathbf{x}_0) \). 
Suppose we have a sequence of latent variables \( \mathbf{x}_t \) for \( t = 1, \ldots, T \), existing within the same sample space as \( \mathbf{x}_0 \), which is denoted as \( \mathcal{X} \). 
DDPMs are latent variable models that are composed of two primary processes: the forward process and the reverse process. 
The forward process is defined by a Markov chain, described as follows:
\begin{equation}
\begin{split}
    &q(\mathbf{x}_{1:T} | \mathbf{x}_0) := \prod_{t=1}^{T} q(\mathbf{x}_t | \mathbf{x}_{t-1}),  \\
    \text{where } &q(\mathbf{x}_t | \mathbf{x}_{t-1}) := \mathcal{N}(\sqrt{1 - \beta_t}\mathbf{x}_{t-1},  \beta_t\mathbf{I}),
\end{split}
\end{equation}
and the variable \(\beta_t\) is a small positive constant indicative of a noise level. The sampling of \(x_t\) can be described by the closed-form expression \(q(x_t | x_0) = \mathcal{N}(x_t; \sqrt{\alpha_t}x_0, (1 - \alpha_t)\mathbf{I})\), where \(\hat{\alpha}_t := 1 - \beta_t\) and \(\alpha_t\) is the cumulative product \(\alpha_t := \prod_{i=1}^t \hat{\alpha}_i\). Consequently, \(x_t\) is given by the equation \(x_t = \sqrt{\alpha_t}x_0 + (1 - \alpha_t)\epsilon\), with \(\epsilon \sim \mathcal{N}(0, \mathbf{I})\). In contrast, the reverse process aims to denoise \(x_t\) to retrieve \(x_0\), a process which is characterized by the ensuing Markov chain:
\begin{equation}
\begin{aligned}
&p_\theta(\mathbf{x}_{0:T}) := p(\mathbf{x}_T) \prod_{t=1}^{T} p_\theta(\mathbf{x}_{t-1} | \mathbf{x}_t), \quad \mathbf{x}_T \sim \mathcal{N}(0, \mathbf{I}), \\
&p_\theta(\mathbf{x}_{t-1} | \mathbf{x}_t) := \mathcal{N}(\mathbf{x}_{t-1}; \mu_\theta(\mathbf{x}_t, t), \sigma^2_\theta(\mathbf{x}_t, t)\mathbf{I}),
\end{aligned}
\end{equation}

\begin{equation}
\begin{aligned}
& \mu_\theta(\mathbf{x}_t, t) = \frac{1}{\alpha_t} \left( \mathbf{x}_t - \frac{\beta_t}{\sqrt{1-\alpha_t}} \epsilon_\theta(\mathbf{x}_t, t) \right), \\
& \sigma_\theta(\mathbf{x}_t, t) = \beta_t^{1/2}, \\
& \text{where } \beta_t = 
\begin{cases} 
\frac{1-\bar{\alpha}_{t-1}}{1-\bar{\alpha}_t} \beta_1, & \text{for } t > 1, \\
\beta_1, & \text{for } t = 1,
\end{cases} \\
& \text{and } \epsilon_\theta(\mathbf{x}_t, t) \text{ is a trainable denoising function}
\end{aligned}
\end{equation}

But in the score-based model, it is slightly different from DDPM. Its forward process is based on a forward stochastic differential equation (SDE), $x(t)$ with $t \in [0,T]$, defined as:
\begin{equation}
    \text{d}x(t) = f(x(t), t)\text{d}t + g(t)\text{d}w,
    \label{sde}
\end{equation}
where $w$ is the standard Wiener process (Brownian motion), $f(\cdot, t): \mathbb R^d \rightarrow \mathbb R^d$ is a vector-valued function called the drift coefficient of $x(t)$. The reverse process is performed via the reverse SDE,
first sampling data $x_t$ from $p(x_t)$ and the generate $x_0$ through the reverse of Eq. (\ref{sde}) as:
\begin{equation}
    \text{d}x(t) = [f(x(t), t) - g(t)^2 \nabla_x \text{log} p(x_t)]\text{d}t + g(t)\text{d} \bar w,
    \label{r_sde}
\end{equation}
where $\bar w$ is the standard Wiener process when time flows backwards from $T$ to 0, and d$t$ is an infinitesimal negative timestep.

The biggest difference between DDPM and score-based model is the estimation of noise equation.
In the DDPM and score-based models, the estimation of the noise function is defined as $\epsilon_{\theta} \left( x_{t},t\right) \ \text{and}~ s_{\theta} \left(x_{t}, t\right)$:
\begin{equation}
    \begin{split}
        s_{\theta} \left(x_{t}, t\right) &= \nabla_{x}\log p_t(x) = - \frac {x_{t} - \sqrt{ \alpha_{t}} x_{0}} {1 -  \alpha_{t}}, \\
    \epsilon_{\theta} \left( x_{t},t\right) &= \frac{x_{t} - \sqrt{ \alpha_{t}} x_{0}} {\sqrt{1 -  \alpha_{t}}},      \\
    s_{\theta} \left(x_{t}, t\right) &= - \frac {1} {\sqrt{1 -  \alpha_{t}}} \epsilon_{\theta} \left( x_{t},t\right),
    \end{split}
    \label{ddpm_convert}
\end{equation}
So we get how to convert DDPM to Score-based model in the sampling stage: $s_{\theta} \left(x_{t}, t\right) = - \frac {1} {\sqrt{1 - \bar \alpha_{t}}} \epsilon_{\theta} \left( x_{t},t\right)$. 

In the DDPM \cite{ddpm2020}, considering the following specific parameterization of $p_\theta(\mathbf{x}_{t-1} | \mathbf{x}_t)$ are learned from the data by optimizing a variational lower bound:
\begin{equation}
\begin{split}
&\log q(x_0) \geq \mathbb{E}_{q(x_0)} [ \underbrace{\log p_{\theta}(x_{0}|x_1)}_{\mathcal{L}_0}  - \underbrace{KL \left( q(x_T|x_0) || q(x_T) \right)}_{\mathcal{L}_T} \\
& -\sum_{t=2}^{T} \underbrace{KL \left( q(x_{t-1} | x_{t}, x_0) || p_{\theta}(x_{t-1} | x_{t}) \right)}_{\mathcal{L}_t} ],\\
\end{split}
\label{vlb}
\end{equation}
where $\mathcal{L}_0$ is that given slightly noisy data $x_1$, correctly reconstruct the original data probability of $x_0$,  $\mathcal{L}_T$ measures the discrepancy between the model distribution $q(x_T|x_0)$ and the prior noise distribution $q(x_T)$ at the last noise level $T$, $\mathcal{L}_t$ measures how the model handles the denoising process from $x_{t}$ to $x_{t-1}$ at each time step, encouraging the model to learn how to remove noise at each step.
In practice, Eq. (\ref{vlb}) can be simplified to the following form:
\begin{equation}\label{ddpm_loss}
\min_{\theta} \mathcal{L}(\theta) := \mathbb{E}_{q(\mathbf{x}_0), \epsilon\sim\mathcal{N}(0,\mathbf{I})}\|\epsilon - \epsilon_{\theta}(\mathbf{x}_t, t)\|^2_2 ,
\end{equation}
where $\mathbf{x}_t = \sqrt{\alpha_t}\mathbf{x}_0 + (1 - \alpha_t)\epsilon$.
The denoising function $\epsilon_\theta$ is designed to estimate the noise vectior $\epsilon$ originally added to the noisy input $x_t$.
Additionally, this training objective also be interpreted as a weighted combination of denoising and score matching, which are techniques commonly employed in the training of score-based generative models. 
\subsection{Forward diffusion process}
Given a data point sampled from a real data distribution $x_o \sim q(x)$, let us define a \textit{forward diffusion process} in which we add small amount of Gaussian noise to the sample in $T$ steps, producing a sequence of noisy samples $x_1, \dots , x_T$. The step sizes are controlled by a variance schedule $\{ \beta_t \in (0, 1) \}_{t=1}^T$.
\begin{equation}
\begin{split}
   &q(x_t | x_{t-1}) = \mathcal{N}(x_t; \sqrt{1 - \beta_{t-1}} x_{t-1}, \beta_{t-1} \mathbf{I}) \\
   &q(x_{1:T} | x_0) = \prod_{t=1}^T q(x_t | x_{t-1})
\end{split}
\end{equation}
The data sample $x_0$ gradually loss its distinguishable features as the step $t$ becomes larger.
Eventually when $T \rightarrow \infty, \ x_{T}$ is equivalent to an isotropic Gaussian distribution.
A nice property of the above process is that we can sample $x_t$ at any arbitrary time step $t$ in a closed form using reparameterization trick.
Let $\alpha_t = 1 - \beta_t$ and $\bar{\alpha}_t = \prod_{i=1}^t \alpha_i$:
\begin{equation}
\begin{split}
x_t &= \sqrt{\alpha_t}x_{t-1} + \sqrt{1 - \alpha_t}\epsilon_{t-1} \\
    &= \sqrt{\alpha_t\alpha_{t-1}}x_{t-2} + \sqrt{1 - \alpha_t\alpha_{t-1}}\bar\epsilon_{t-2} \\
    & \quad \vdots \\
    &= \sqrt{\overline{\alpha}_t}x_0 + \sqrt{1 - \overline{\alpha}_t}\epsilon \\
q(x_t|x_0) &= \mathcal{N}(x_t; \sqrt{\overline{\alpha}_t}x_0, (1 - \overline{\alpha}_t)I)
\end{split}
\end{equation}
where$\epsilon_{t-1}, \epsilon_{t-2}, \ldots \sim \mathcal{N}(0,I)$; where$\bar \epsilon_{t-2}$merges two Gaussians(*).
(*) Recall that when we merge two Gaussians with different variance, $\mathcal{N}(0, \sigma_1^2\text{I})$ and $\mathcal{N}(0, \sigma_2^2\text{I})$,
the new distribution is $\mathcal{N}(0, (\sigma_1^2 + \sigma_2^2)\text{I})$.
 Here the merged standard deviation is $\sqrt{(1 - \alpha_t)+\alpha_t(1 - \alpha_{t-1})} = \sqrt{1- \alpha_t \alpha_{t-1}}$.
 Usually, we can afford a larger update step when the sample gets noisier, so $\beta_1 < \beta_2 < \ldots < \beta_T$ and therefore $\bar \alpha_1 > \ldots > \bar \alpha_T$

\subsection{Connection with stochastic gradient Langevin dynamics in score-base model}
Langevin dynamics is a concept from physics, developed for statistically modeling molecular systems. Combined with stochastic gradient descent, stochastic gradient Langevin dynamics can produce samples from a probability density $p(x)$ using only the gradients $\nabla_x \text{log} p(x)$ in a Markov chain of updates:
\begin{equation}
x_t = x_{t-1} + \frac{\delta}{2} \nabla_{x} \log p(x_{t-1}) + \sqrt{\delta}\epsilon_t, \quad \text{where } \epsilon_t \sim \mathcal{N}(0, I)
\end{equation}
where $\delta$ is the stpe size. When $T \rightarrow \infty, \epsilon \rightarrow 0, x_T$ equals to the true probability density $p(x)$.
Compared to standard SGD, stochastic gradient Langevin dynamics injects Gaussian noise into the parameter updates to avoid collapses into local minima.

\subsection{Reverse diffusion process}
If we can reverse the above process and sample from $q(x_{t-1} | x_t)$, we we will be able to recreate the true sample from a Gaussian noise input, $x_T \sim \mathcal{N}(0, \text{I})$. 
Note that if $\beta_t$ is small enough, $q(x_{t-1}|x_t)$ will also be Gaussian.
Unfortunately, we cannot easily estimate $q(x_{t-1}|x_t)$ because it needs to use the entire dataset and therefore we need to learn a model $p_\theta$ to approximate these conditional probabilities in order to run the reverse diffusion process.
\begin{equation}
\begin{split}
    &p_{\theta}(x_0:T) = p_{\theta}(x_T) \prod_{t=1}^{T} p_{\theta}(x_{t-1}|x_t) \\
    &p_{\theta}(x_{t-1}|x_t) = \mathcal{N}(x_{t-1}; \mu_{\theta}(x_t, t), \Sigma_{\theta}(x_t, t))
\end{split}
\end{equation}
It is noteworthy that the reverse conditional probability is tractable when conditioned on $x_0$:
\begin{equation}
q(x_{t-1} | x_t, x_0) = \mathcal{N}(x_{t-1}; \mu(x_t, x_0), \beta I)
\end{equation}
Using Bayes’ rule, we have:
\begin{equation}
\small
\begin{split}
&q(x_{t-1} | x_t, x_0) = \frac{q(x_{t-1}, x_t)}{q(x_t | x_0)} \\
&\propto \exp\left( -\frac{1}{2} \Bigg( \frac{(x_t - \sqrt{\alpha_t}x_{t-1})^2}{\beta_t} \right. \\
&\quad + \left. \frac{(x_{t-1} - \sqrt{\overline{\alpha}_{t-1}}x_0)^2}{1 - \overline{\alpha}_{t-1}} - \frac{(x_t - \sqrt{\overline{\alpha}_t}x_0)^2}{1 - \overline{\alpha}_t} \Bigg) \right) \\
&= \exp\left( -\frac{1}{2} \Bigg( \frac{x_t^2}{\beta_t} - 2\frac{\sqrt{\alpha_t}x_{t-1}x_t}{\beta_t} \right. \\
&\quad + \left. \frac{\alpha_t x_{t-1}^2}{\beta_t} - 2\frac{\sqrt{\overline{\alpha}_{t-1}} x_{t-1}x_0}{1 - \overline{\alpha}_{t-1}} + \frac{x_{t-1}^2}{1 - \overline{\alpha}_{t-1}} + C(x_t, x_0) \Bigg) \right) \\
&= \exp\left( -\frac{1}{2} \Bigg( \left( \frac{\alpha_t}{\beta_t} + \frac{1}{1 - \overline{\alpha}_{t-1}} \right) x_{t-1}^2 \right. \\
&\quad - 2\left( \frac{\sqrt{\alpha_t}}{\beta_t} + \frac{\sqrt{\overline{\alpha}_{t-1}}}{1 - \overline{\alpha}_{t-1}} \right) x_{t-1}x_t \\
&\quad \left. + C(x_t, x_0) \Bigg) \right)
\end{split}
\end{equation}

where $C(x_t, x_0)$ is some function not involving $x_{t-1}$ and details are omitted. Following the standard Gaussian density function, the mean and variance can be parameterized as follows (recall that $\alpha_t = 1 - \beta_t$ and $\bar{\alpha}_t = \prod_{i=1}^t \alpha_i$):
\begin{equation}
\begin{split}
\hat{\beta}_t &= \frac{1}{\left( \frac{\alpha_t}{\beta_t} + \frac{1}{1 - \overline{\alpha}_{t-1}} \right)} = \frac{1}{\left( \frac{\alpha_t - \overline{\alpha}_t + \beta_t}{\beta_t(1 - \overline{\alpha}_{t-1})} \right)} \\ 
&= \frac{1 - \overline{\alpha}_{t-1}}{\alpha_t - \overline{\alpha}_t + \beta_t} 
= \frac{1 - \overline{\alpha}_{t-1}}{1 - \overline{\alpha}_t} \cdot \beta_t, \\
\hat{\mu}(x_t, x_0) &= \left( \frac{\sqrt{\overline{\alpha}_t} x_t}{\beta_t} + \frac{\sqrt{1 - \overline{\alpha}_{t-1}} x_0}{1 - \overline{\alpha}_{t-1}} \right) \left/ \left( \frac{\alpha_t}{\beta_t} + \frac{1}{1 - \overline{\alpha}_{t-1}} \right) \right. \\
&= \left( \frac{\sqrt{\overline{\alpha}_t} x_t}{\beta_t} + \frac{\sqrt{1 - \overline{\alpha}_{t-1}} x_0}{1 - \overline{\alpha}_{t-1}} \right) \cdot \frac{1 - \overline{\alpha}_{t-1}}{1 - \overline{\alpha}_t} \cdot \beta_t \\
&= \frac{\sqrt{\overline{\alpha}_t (1 - \overline{\alpha}_{t-1})} x_t + \sqrt{\overline{\alpha}_{t-1} \beta_t} x_0}{1 - \overline{\alpha}_t}.
\end{split}
\end{equation}
We can represent $x_0 = \frac{1}{\sqrt{\overline{\alpha}_t}} \left( x_t - \sqrt{1 - \overline{\alpha}_t} \epsilon_t \right)$ and plug into the above equation and obtain:
\begin{equation}
\begin{split}
\hat \mu_t &= \frac{\sqrt{\overline{\alpha}_t (1 - \overline{\alpha}_{t-1})}}{1 - \overline{\alpha}_t} x_t + \frac{\sqrt{\overline{\alpha}_{t-1} \beta_t}}{1 - \overline{\alpha}_t} \frac{1}{\sqrt{\overline{\alpha}_t}} \left( x_t - \sqrt{1 - \overline{\alpha}_t} \epsilon_t \right) \\
&= \frac{1}{\sqrt{{\alpha}_t}} \left( x_t - \frac{1 - {\alpha}_t}{\sqrt{1 - \overline{\alpha}_t}} \epsilon_t \right)
\end{split}
\end{equation}
such a setup is very similar to VAE and thus we can use the variational lower bound to optimize the negative log-likelihood.
\begin{equation}
\tiny
\begin{split}
-\log p_{\theta}(x_0) &\leq -\log p_{\theta}(x_0) + D_{KL}(q(x_{1:T}|x_0) \| p_{\theta}(x_{1:T}|x_0)) \\
&= -\log p_{\theta}(x_0) + \mathbb{E}_{x_{1:T} \sim q(x_{1:T}|x_0)} \left[ \log \frac{q(x_{1:T}|x_0)}{p_{\theta}(x_{0:T})\ / p_{\theta}(x_0)} \right] \\
&= -\log p_{\theta}(x_0) + \mathbb{E}_{q} \left[ \log \frac{q(x_{1:T}|x_0)}{p_{\theta}(x_{0:T})} + \log p_{\theta}(x_0) \right] \\
&= \mathbb{E}_{q} \left[ \log \frac{q(x_{1:T}|x_0)}{p_{\theta}(x_{0:T})} \right] \\
\text{Let } L_{VLB} &= \mathbb{E}_{q(x_{0:T})} \left[ \log \frac{q(x_{1:T}|x_0)}{p_{\theta}(x_{0:T})} \right] \geq -\mathbb{E}_{q(x_0)} \log p_{\theta}(x_0)
\end{split}
\end{equation}
It is also straightforward to get the same result using Jensen’s inequality. Say we want to minimize the cross entropy as the learning objective,
\begin{equation}
\begin{split}
L_{CE} &= -\mathbb{E}_{q(x_0)} \log p_{\theta}(x_0) \\
&= -\mathbb{E}_{q(x_0)} \log \left( \int p_{\theta}(x_0:T) \, dx_1:T \right) \\
&= -\mathbb{E}_{q(x_0)} \log \left( \int \frac{q(x_1:T|x_0) p_{\theta}(x_0:T)}{q(x_1:T|x_0)} \, dx_1:T \right) \\
&= -\mathbb{E}_{q(x_0)} \log \left( \mathbb{E}_{q(x_1:T|x_0)} \left[ \frac{p_{\theta}(x_0:T)}{q(x_1:T|x_0)} \right] \right) \\
&\leq -\mathbb{E}_{q(x_0:T)} \log \frac{p_{\theta}(x_0:T)}{q(x_1:T|x_0)} \\
&= \mathbb{E}_{q(x_0:T)} \log \frac{q(x_1:T|x_0)}{p_{\theta}(x_0:T)} \\
&= L_{VLB}
\end{split}
\end{equation}
To convert each term in the equation to be analytically computable, the objective can be further rewritten to be a combination of several KL-divergence and entropy terms:
\begin{equation}
\tiny
\begin{aligned}
& L_{VLB} = \mathbb{E}_{q(x_0:T)} \left[ \log \frac{q(x_1:T|x_0)}{p_{\theta}(x_0:T)} \right] \\
&= \mathbb{E}_{q} \left[ \log \frac{q(x_T|x_0)}{p_{\theta}(x_T)} \prod_{t=1}^{T} \frac{q(x_{t-1}|x_t)}{p_{\theta}(x_{t-1}|x_t)} \right] \\
&= \mathbb{E}_{q} \left[ -\log p_{\theta}(x_T) + \sum_{t=1}^{T} \log \frac{q(x_{t-1}|x_t)}{p_{\theta}(x_{t-1}|x_t)} \right] \\
&= \mathbb{E}_{q} \left[ -\log p_{\theta}(x_T) + \log \frac{q(x_{T-1}|x_T)}{p_{\theta}(x_{T-1}|x_T)}  + \log \frac{q(x_1|x_2)}{p_{\theta}(x_1|x_2)} \right] \\
&= \mathbb{E}_{q} \left[ -\log p_{\theta}(x_T) + \sum_{t=2}^{T} \log \frac{q(x_{t-1}|x_t, x_0)}{p_{\theta}(x_{t-1}|x_t)} + \log \frac{q(x_1|x_0)}{p_{\theta}(x_1|x_0)} \right] \\
&= \mathbb{E}_{q} \left[ -\log p_{\theta}(x_T) + \sum_{t=2}^{T} \log \frac{q(x_{t-1}|x_t, x_0)}{p_{\theta}(x_{t-1}|x_t)} \right] + \mathbb{E}_{q} \left[ \log \frac{q(x_1|x_0)}{p_{\theta}(x_1|x_0)} \right] \\
&= \mathbb{E}_{q} \left[ -\log p_{\theta}(x_T) + \sum_{t=2}^{T} \log \frac{q(x_{t-1}|x_t, x_0)}{p_{\theta}(x_{t-1}|x_t)} \right] - \mathbb{E}_{q(x_0)} \log p_{\theta}(x_0|x_1) \\
&= \mathbb{E}_{q} \underbrace{\left[ \log \frac{q(x_T|x_0)}{p_{\theta}(x_T)} \right]}_{L_T} + \sum_{t=2}^{T} \mathbb{E}_{q} \underbrace{\left[ D_{KL}(q(x_{t-1}|x_t, x_0) || p_{\theta}(x_{t-1}|x_t)) \right]}_{L_{t-1}}- \\
&~~~~~~ \mathbb{E}_{q(x_0)} \underbrace{\log p_{\theta}(x_0|x_1)}_{L_0}
\end{aligned}
\end{equation}
Let’s label each component in the variational lower bound loss separately:
\begin{equation}
\begin{aligned}
L_{VLB} &= L_T + L_{T-1} + \cdots + L_0 \\
\text{where} \quad L_T &= D_{KL}(q(x_T|x_0) \| p_{\theta}(x_T)), \\
L_t &= D_{KL}(q(x_{t}|x_{t+1}, x_0) \| p_{\theta}(x_{t}|x_{t+1})) \quad \\
& \text{for } 1 < t < T, \\
L_0 &= -\log p_{\theta}(x_0|x_1).
\end{aligned}
\end{equation}
Every KL term in $L_{VLB}$ (except for $L_0$) compares two Gaussian distributions and therefore they can be computed in closed form. 
$L_T$ is constant and can be ignored during training because $q$ has no learnable parameters and $x_T$is a Gaussian noise.
The DDPM models $L_0$ using a separate discrete decoder derived from $\mathcal{N}(\mathbf{x_0}; \mu_{\theta}(\mathbf{x_1}, 1), \Sigma_{\theta}(\mathbf{x_1}, 1)).$

\subsection{Parameterization of $L_t$ for Training Loss}
Recall that we need to learn a neural network to approximate the conditioned probability distributions in the reverse diffusion process, $p_{\theta}(x_{t-1}|x_t) = \mathcal{N}(\mathbf{x_{t-1}}; \mu_{\theta}(\mathbf{x_t}, t), \Sigma_{\theta}(\mathbf{x_t}, t)).$
We would like to train $\mu_\theta$ to predict $\hat \mu_t = \frac{1}{\sqrt{{\alpha}_t}} \left( x_t - \frac{1 - {\alpha}_t}{\sqrt{1 - \overline{\alpha}_t}} \epsilon_t \right).$
Because $x_t$ is available as input at training time, we can reparameterize the Gaussian noise term instead to make it predict $\epsilon_t$ from the input $x_t$ at time step $t$:
\begin{equation}
\small
\begin{split}
\mu_{\theta}(x_t, t) &= \frac{1}{\sqrt{\alpha_t}} \left( x_t - \frac{1 - \alpha_t}{\sqrt{1 - \alpha_t}} \epsilon_{\theta}(x_t, t) \right) \\
\text{Thus } x_{t-1} &\sim \mathcal{N}\left(x_{t-1}; \frac{1}{\sqrt{\alpha_t}} \left( x_t - \frac{1 - \alpha_t}{\sqrt{1 - \alpha_t}} \epsilon_{\theta}(x_t, t) \right), \Sigma_{\theta}(x_t, t)\right)
\end{split}
\end{equation}
The loss term $L_t$ is parameterized to minimize the difference from $\hat \mu$:
\begin{equation}
\begin{split}
L_t &= \mathbb{E}_{x_0,\epsilon_t} \left[ \frac{1}{2} \frac{\| \mu_{\theta}(x_t, x_0) - \mu_{\theta}(x_t, t) \|^2}{\| \Sigma_{\theta}(x_t, t) \|_2^2} \right] \\
&= \mathbb{E}_{x_0,\epsilon_t} \Bigg[ \frac{1}{2} \frac{1}{\| \Sigma_{\theta}(x_t, t) \|_2^2} \Bigg\| \frac{1}{\sqrt{\alpha_t}} (x_t - \frac{1 - \alpha_t}{\sqrt{1 - \alpha_t}} \epsilon_t) \\
&\quad - \frac{1}{\sqrt{\alpha_t}} (x_t - \frac{1 - \alpha_t}{\sqrt{1 - \alpha_t}} \epsilon_{\theta}(x_t, t)) \Bigg\|^2 \Bigg] \\
&= \mathbb{E}_{x_0,\epsilon_t} \left[ \frac{1}{2\alpha_t(1 - \alpha_t)} \frac{1}{\| \Sigma_{\theta}(x_t, t) \|_2^2} \| \epsilon_t - \epsilon_{\theta}(x_t, t) \|^2 \right] \\
&= \mathbb{E}_{x_0,\epsilon_t} \Bigg[ \frac{1}{2\alpha_t(1 - \alpha_t)} \frac{1}{\| \Sigma_{\theta}(x_t, t) \|_2^2} \\
&\quad \times \| \epsilon_t - \epsilon_{\theta}(\sqrt{\alpha_t} x_0 + \sqrt{1 - \alpha_t} \epsilon_t, t) \|^2 \Bigg]
\end{split}
\end{equation}

\subsection{Connection with noise-conditioned score networks (NCSN) in score-based model}
Song \& Ermon (2019) proposed a score-based generative modeling method where samples are produced via \textit{Langevin dynamics} using gradients of the data distribution estimated with score matching. The score of each sample \( \mathbf{x} \)'s density probability is defined as its gradient \( \nabla_{\mathbf{x}} \log q(\mathbf{x}) \). A score network \( s_{\theta} : \mathbb{R}^D \rightarrow \mathbb{R}^D \) is trained to estimate it, \( s_{\theta}(\mathbf{x}) \approx \nabla_{\mathbf{x}} \log q(\mathbf{x}) \).

To make it scalable with high-dimensional data in the deep learning setting, they proposed to use either \textit{denoising score matching} (Vincent, 2011) or \textit{sliced score matching} (use random projections; Song et al., 2019). Denoising score matching adds a pre-specified small noise to the data \( \tilde{q}(\mathbf{x}) \) and estimates \( q(\mathbf{x}) \) with score matching.

Recall that Langevin dynamics can sample data points from a probability density distribution using only the score \( \nabla_{\mathbf{x}} \log q(\mathbf{x}) \) in an iterative process.

However, according to the manifold hypothesis, most of the data is expected to concentrate in a low dimensional manifold, even though the observed data might look only arbitrarily high-dimensional. It brings a negative effect on score estimation since the data points cannot cover the whole space. In regions where data density is low, the score estimation is less reliable. After adding a small Gaussian noise to make the perturbed data distribution cover the full space \( \mathbb{R}^D \), the training of the score estimator network becomes more stable. Song \& Ermon (2019) improved it by perturbing the data with the noise of \textit{different levels} and train a noise-conditioned score network to \textit{jointly} estimate the scores of all the perturbed data at different noise levels.

The schedule of increasing noise levels resembles the forward diffusion process. If we use the diffusion process annotation, the score approximates \( s_{\theta}(x_t, t) \approx \nabla_{\mathbf{x}} \log q(\mathbf{x}) \). Given a Gaussian distribution \( \mathcal{N}(\mu, \sigma^2) \), we can write the derivative of the logarithm of its density function as
$\nabla_{\mathbf{x}} \log p(\mathbf{x}) = \nabla_{\mathbf{x}} \left( -\frac{(\mathbf{x} - \mu)^2}{2\sigma^2} \right) = -\frac{\mathbf{x} - \mu}{\sigma^2} = -\frac{\mathbf{x} - \mu}{\sigma^2} - \mathbf{\epsilon}$ where $ \mathbf{\epsilon} \sim \mathcal{N}(0, I).$Recall that$q(\mathbf{x}_t) \sim \mathcal{N}\left(\sqrt{\alpha_t}\mathbf{x}_0, (1 - \alpha_t)I\right)$
and therefore,
\begin{equation}
\begin{split}
    s_{\theta}(\mathbf{x}_t, t) &\approx \nabla_{\mathbf{x}} \log q(\mathbf{x}) = \mathbb{E}_{q(\mathbf{x}_0)} \left[ \nabla_{\mathbf{x}} \log q(\mathbf{x}_t | \mathbf{x}_0) \right] \\
    &= \mathbb{E}_{q(\mathbf{x}_0)} \left[ -\frac{\mathbf{\epsilon}_t(\mathbf{x}_t, t)}{\sqrt{1 - \alpha_t}} \right] = -\frac{\mathbf{\epsilon}_t(\mathbf{x}_t, t)}{\sqrt{1 - \alpha_t}}.
\end{split}
\end{equation}

\section{stMCDI algorithm} \label{ap_algorithm}
We provide the training procedure of stMCDI in Algorithm 1 and the imputation (sampling) procedure with stMCDI in Algorithm 2.
It should be noted that each of our samples is a spot, written as $x_{j, t}$ in the algorithm, where $j= 1, \dots,n$, $t$ represents the time step in diffusion.
\begin{algorithm}
    \caption{Training of stMCDI}
    \SetAlgoLined
     \LinesNumbered
    \KwIn{ST data $X = \{x_j\}_{j=1}^n$, Adjacency matrices $A$, Number of iteration \( N_{\text{iter}} \),  \( \{\alpha_t\}_{t=1}^T \), $T$ .}
    \KwOut{Trained denoising function \( s_\theta \)}

    \For {\( i = 1 \) to \( N_{\text{iter}} \)} {
         $x_j \sim X $,\\
         $\hat{x}_j = GCN(x_j, A)$,\\
         \( t \sim \text{Uniform}(\{1, \ldots, T\})  \),\\
         Separate observed values of \( \hat{x}_{j, 0} \) into conditional information \( \hat{x}_{j, 0}^c \) and imputation targets \(  \hat{x}_{j, 0}^{*} \),\\
         \( \epsilon \sim \mathcal{N}(0, I) \),\\
         Take gradient step on \\
    \( \nabla_{\theta} \|(\epsilon - s_{\theta}(\sqrt{\alpha_t} \hat{x}_{j, 0}^{*} + \sqrt{(1-\alpha_t)} \epsilon, t | \hat{x}_{j, 0}^c)) \odot M\|^2_2 \),
    }
\end{algorithm}


\begin{algorithm}
    \caption{Training of stMCDI}
    \SetAlgoLined
    \LinesNumbered
    \KwIn{Data sample \( x_{j, 0} \), Denoising function \( s_\theta \)}
    \KwOut{Imputed missing values \( x_{j, 0}^{*} \)}

    \For{\( t = T \) to 1}{
        Sample $\epsilon_t \sim \mathcal{N}(0, I)$;\\
        $x_{j, t-1} = \frac{1}{\sqrt{\alpha_t}} \left( x_{j,t} - \frac{1-\alpha_t}{\sqrt{1-\bar{\alpha}_t}} s_{\theta}(x_{j,t}, t) \right) + \sqrt{\beta_t} \epsilon_t$,
        $t = t - 1$,
    }
\end{algorithm}

\section{Details of experiment settings} \label{ap_hyper}
In this section, we provide the details of the experiment settings in our model.
When we evaluate baseline methods, we use their original hyperparameters and model sizes.
As for hyperparameters, we set the batch size as 64 and the number of epochs as 2000.
We used Adam optimizer with learning rate 6e-6 that is decayed to 3e-7 and 6e-9 at 50\% and 75\% of the total epochs, respectively.
As for the model, we set the number of residual layers as 4, residual channels as 32, and attention heads as 16.
We followed \cite{diffwave} for the number of channels and decided the number of layers based on the validation loss and the parameter size.
The number of the parameter in the model is about 826,471.

We also provide hyperparameters for the diffusion model as follows.
We set the number of the diffusion step $T = 2000$, the minimum noise level $\beta_1 = 10e-6$, and the maximum noise level $\beta_T = 0.05$.
Since recent studies \cite{ddim,iddpm} reported that gentle decay of $\alpha_t$ could improve the sample quality, we adopted the following quadratic schedule for other noise levels:
\begin{equation}
    \beta_t = \left( \frac{T-t}{T-1} \sqrt{\beta_1} + \frac{t-1}{T-1} \sqrt{\beta_T} \right)^2 
\end{equation}
\section{Details of Metrics} \label{ap_metrics}
To evaluate the performance of stMCDI, we use four key evaluation metrics: Pearson Correlation Coefficient (PCC), Cosine Similarity (CS), Root Mean Square Error (RMSE), and Mean Absolute Error (MAE). 
Given the nature of spatial transcriptomic data, which lacks real labels, we utilize the masked part values as surrogate ground truths. 
Consequently, the calculation of these four evaluation metrics is confined exclusively to the masked portions, providing a standardized basis for gauging model performance.

The definitions of the four evaluation indicators are as follows:
\begin{equation}
PCC = \frac{\sum{(x_i - \bar{x})(y_i - \bar{y})}}{\sqrt{\sum{(x_i - \bar{x})^2}\sum{(y_i - \bar{y})^2}}},
\end{equation}
\begin{equation}
CS = \frac{\sum_{i=1}^{n} x_i y_i}{\sqrt{\sum_{i=1}^{n} x_i^2} \times \sqrt{\sum_{i=1}^{n} y_i^2}},
\end{equation}
\begin{equation}
RMSE = \sqrt{\frac{1}{n} \sum_{i=1}^{n} (x_i - y_i)^2},
\end{equation}
\begin{equation}
MAE = \frac{1}{n} \sum_{i=1}^{n} |x_i - y_i|,
\end{equation}
where $x_i$ means the prediction values and $\bar x$ is its mean values, $y_i$ is the mask part of the data and $\bar y$ s its mean values.

\section{Details of datasets} \label{ap_data} 
\subsection{Data sources and data preprocessing}
We compared the performance of our model with other baseline methods on 6 real-world spatial transcriptomic datasets from several representative sequencing platforms. The 6 real-world spatial transcriptomics datasets used in our experiments were derived from recently published papers on spatial transcriptomics experiments, and details are shown in \cite{stBenchmark}.
All six datasets come from different species, including mice and humans, and different organs, such as brain, lungs and kidneys. Specifically, the number of spots ranges from 278 to 6000, and the gene range ranges from 14192 to 28601.
We preprocess each dataset through the following steps. (1) Normalization of expression matrix. For spatial transcriptomics datasets, we tested unnormalized and normalized expression matrices input to each integration method. To normalize the expression matrix, we use the following equation:
\begin{equation}
    D_{ij} = \log ( N \times \frac{C_{ij}} {\sum\limits_{j=1}^{M} C_{ij}} + 1  )
\end{equation}
where $C_{ij}$ represents the raw read count of gene $i$ at spot $j$, $D_{ij}$ represents the normalized read count of gene $i$ at spot $j$, $N$ is the median number of transcripts detected per spot. 
(2)Selection of highly variable genes. For spatial transcriptome datasets with more than 1,000 detected genes, we calculated the coefficient of variation of each gene using the following equation:
\begin{equation}
    CV_i= \frac{\sigma_i}{u_i}
\end{equation}
where $CV_i$ is the coefficient of variation of gene $i$; $\sigma_i$ is s.d. of the spatial distribution of gene $i$ at all spots; $u_i$ is the average expression of gene $i$ at all sites. 
We used this method to screen and obtain 1,000 gene features.
Specific details for each dataset are shown in the Table. \ref{dataset}

\begin{table*}[htbp]
\centering
\resizebox{\textwidth}{!}{%
\begin{tabular}{c|ccccc}
\hline
\textbf{Dataset} & \textbf{Tissue Section}  & \textbf{GEO ID} & \textbf{Number of spots} & \textbf{Download Link}                                             \\ \hline
MOB              & Mouse Olfactory Bulb                      & GSE121891       & 6225                     & \url{https://www.ncbi.nlm.nih.gov/geo/query/acc.cgi?acc=GSE121891} \\
HBC              & Human Breast Cancer                        & GSE176078       & 4785                     & \url{https://www.ncbi.nlm.nih.gov/geo/query/acc.cgi?acc=GSE176078} \\
HP               & Human Prostate                             & GSE142489       & 2278                     & \url{https://www.ncbi.nlm.nih.gov/geo/query/acc.cgi?acc=GSE142489} \\
HO               & Human Osteosarcoma                        & GSE152048       & 1646                     & \url{https://www.ncbi.nlm.nih.gov/geo/query/acc.cgi?acc=GSE152048} \\
ML               & Mouse Liver                                  & GSE109774       & 2178                     & \url{https://www.ncbi.nlm.nih.gov/geo/query/acc.cgi?acc=GSE109774} \\
MK               & Mouse Kidney                                 & GSE154107       & 1889                     & \url{https://www.ncbi.nlm.nih.gov/geo/query/acc.cgi?acc=GSE154107} \\ \hline
\end{tabular}%

}
\caption{The detail of each datasets.}
\label{dataset}
\end{table*}

\section{Additional examples of stMCDI imputation}
\subsection{Imputation performance of different generative models}
\begin{figure*}[htbp]
    \centering
    \includegraphics[width=0.7\textwidth]{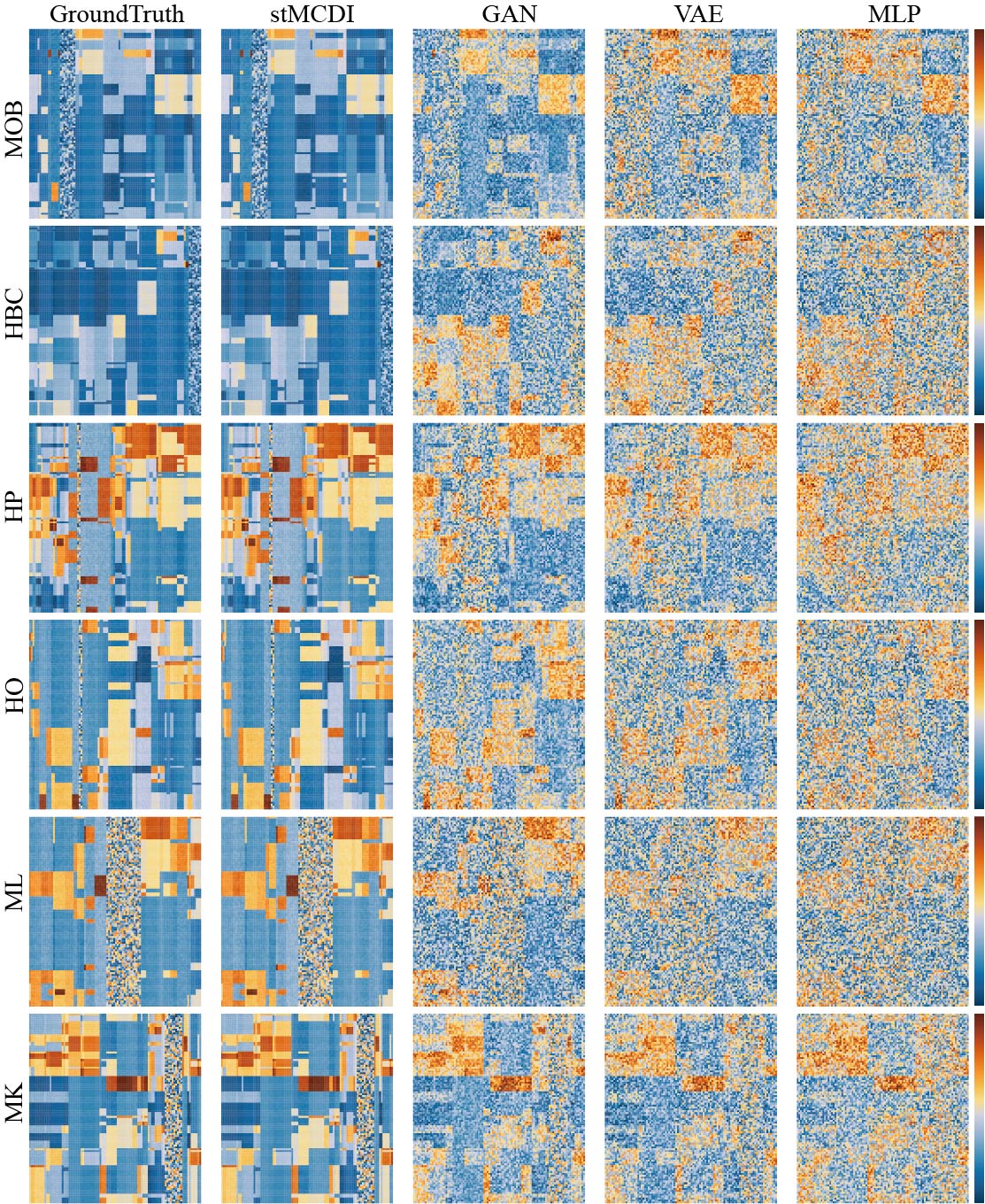}
    \caption{Visualization of the imputation performance of various generative methods across six distinct spatial transcriptomic datasets.}
    \label{fig4}
\end{figure*}

In order to better verify the effectiveness of our proposed method, we also compared stMCDI with several common generative models.
To maintain experimental consistency, we still perform diffusion and reduction on the latent representation constructed by the graph encoder.
As before, we use the mask part data as the real label, and visualize the prediction part of each model through clustering, as shown in the Fig. \ref{fig4}.
We used GAN \cite{GAIN}, VAE \cite{vaei}, and a three-layer MLP.
Experimental results show that our proposed method achieves the best performance in the imputation of missing values.

\subsection{More results in Basline comparison} \label{ap_more_baseline}
Due to the limited length of the text, some baseline comparisons are placed in the supplementary materials, as shown in the Fig. \ref{fig5}.
\begin{figure*}[htbp]
    \centering
    \includegraphics[width=1\textwidth]{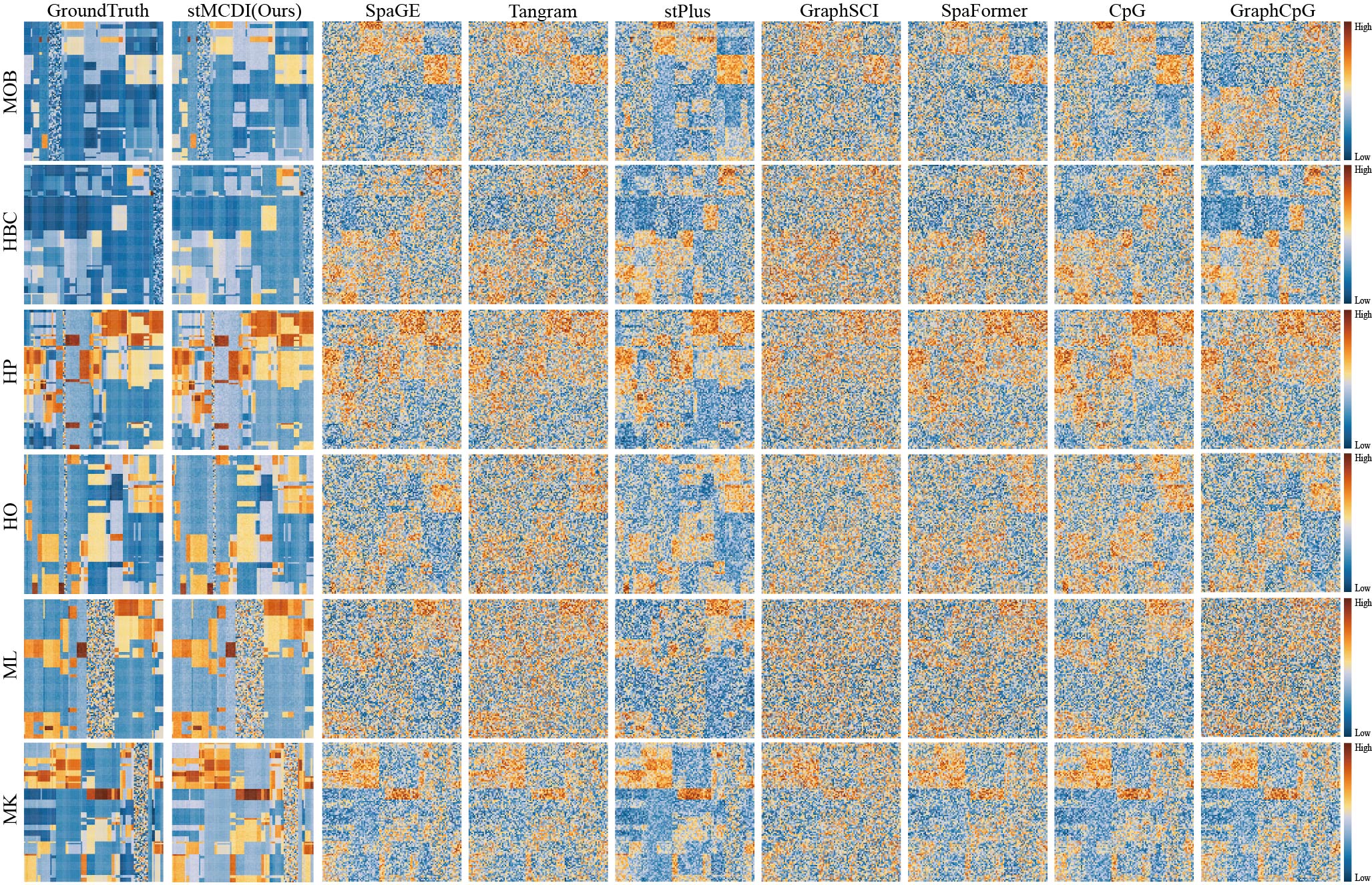}
    \caption{Visualization of the imputation performance of various baseline methods across six distinct spatial transcriptomic datasets.}
    \label{fig5}
\end{figure*}


\end{document}